\documentclass{aa}

\usepackage{txfonts}

\usepackage{graphicx}

\usepackage{natbib}
\bibpunct{(}{)}{;}{a}{}{,} 

\begin{document}

\title{CFBDS J005910.90-011401.3: reaching the T-Y Brown Dwarf transition?
\thanks{Based in part on observations obtained with MegaPrime/MegaCam, a 
joint project of CFHT and CEA/DAPNIA, at the Canada-France-Hawaii 
Telescope (CFHT) which is operated by the National Research Council 
(NRC) of Canada, the Institut National des Sciences de l'Univers of 
the Centre National de la Recherche Scientifique (CNRS) of France, 
and the University of Hawaii. This work is based in part on data 
products produced at TERAPIX and the Canadian Astronomy Data Centre 
as part of the Canada-France-Hawaii Telescope Legacy Survey, a 
collaborative project of NRC and CNRS. Also based  on 
observations obtained at the Gemini Observatory, which is operated 
by the Association of Universities for Research in Astronomy, 
Inc., under a cooperative agreement with the NSF on behalf of 
the Gemini partnership: the National Science Foundation (United
States), the Science and Technology Facilities Council (United 
Kingdom), the National Research Council (Canada), CONICYT (Chile), 
the Australian Research Council (Australia), CNPq (Brazil) and 
SECYT (Argentina) under programs GN-2007A-Q-201 and
GN-2007B-Q-3. Also based on observations made with ESO  
Telescopes at the La Silla Observatory under 
programmes 078.C-0629 and 078.A-0651.
Some of the data presented herein were obtained at the
  W.M. Keck Observatory, which is operated as a scientific partnership
  among the California Institute of Technology, the University of
  California, and the National Aeronautics and Space Administration. The
  Observatory was made possible by the generous financial support of the
  W.M. Keck Foundation. }
 }

\author{
P. Delorme \inst{1}
\and X. Delfosse \inst{1}
\and L. Albert \inst{2}
\and E. Artigau \inst{3}
\and T. Forveille \inst{1,4}
\and C. Reyl\'e \inst{5}
\and F. Allard \inst{6}
\and D. Homeier \inst{7}
\and A.C. Robin \inst{5}
\and C.J. Willott \inst{8}
\and Michael C. Liu  \inst{4} \thanks{Alfred P. Sloan Research Fellow}
\and T.J. Dupuy \inst{4}
}

\offprints{P. Delorme, \email{Philippe.Delorme@obs.ujf-grenoble.fr}}

\institute{Laboratoire d'Astrophysique de Grenoble,
               Observatoire de Grenoble,
               Universit\'e Joseph Fourier,
               CNRS, UMR 571
               Grenoble,
               France
\and
               Canada-France-Hawaii Telescope Corporation,
               65-1238 Mamaloha Highway,
               Kamuela, HI 96743,
               U.S.A.
\and
               Gemini Observatory, 
               Southern Operations Center, 
               Association of Universities for Research in Astronomy, Inc., 
               Casilla 603, La Serena, Chile.
\and  
	       Institute for Astronomy,
               2680 Woodlawn Drive, Honolulu,
               HI 96822-1839, USA.
\and  
               Observatoire de Besan\c{c}on, 
               Institut UTINAM, University of Franche-Comt\'e, 
               CNRS-UMR 6213,
               BP 1615, 25010 Besançon Cedex, France
\and 
               Centre de Recherche Astrophysique de Lyon,  
               UMR 5574: CNRS, Universit\'e de Lyon, 
               Ecole Normale Sup\'erieure de Lyon, 
               46 all\'ee d'Italie, 69364 Lyon Cedex 07, France
\and
               Institut f\"ur Astrophysik, Georg-August-Universit\"at, 
               Friedrich-Hund-Platz 1, 37077 G\"ottingen, Germany
\and
               Physics Department, University of Ottawa, 
               150 Louis Pasteur, MacDonald Hall, Ottawa, ON K1N 6N5, Canada
}

\abstract
{
{
We report the discovery of CFBDS J005910.90-011401.3 (hereafter CFBDS0059), 
the coolest brown dwarf identified to date.}
{ We found CFBDS0059
using $i'$ and $z'$ images from the Canada-France-Hawaii Telescope
(CFHT), and present optical and 
near-infrared photometry,  Keck laser guide star adaptive optics imaging, and
a complete near-infrared spectrum,  from 1.0 to 2.2 $\mu$m. }
{A side to side comparison of the near-infrared spectra of 
CFBDS0059 and ULAS~J003402.77-005206.7 (hereafter ULAS0034), previously 
the coolest known brown dwarf, indicates that CFBDS0059 is 
$\sim50~\pm~15$K cooler.
We estimate a temperature of $T_\mathrm{eff} \sim 620$K and gravity of $\log g
\sim 4.75$. Evolutionary models translate these parameters into an
age of 1-5 Gyr and a mass of $15-30$ M$_{Jup}$. We
estimate a photometric distance of $\sim$13pc, which puts CFBDS0059 
within easy reach of accurate parallax measurements. Its large proper motion 
suggests membership in the older population of the thin disk.
The spectra of both CFBDS0059 and ULAS~J0034 shows probable 
absorption by  a wide ammonia band on the blue side of the 
$H$-band flux peak. If, as we expect, that 
feature deepens further for still lower effective temperatures, its 
appearance will become a natural breakpoint for the transition between
the T spectral class and the new Y spectral type. CFBDS0059 and 
ULAS~J0034 would then be the first Y0 dwarfs.
}{}

\date{}

}

\maketitle

\section{Introduction}

Observed stellar and substellar atmospheres cover a continuum of physical
conditions from the hottest stars ($\sim$~100~000~K) to the coolest 
known brown dwarfs (previously ULAS~J003402.77-005206.7 (hereafter 
ULAS0034), $>$T8
 \citet{warren.2007}). There
remains however a sizeable temperature gap between  the 600-700~K
ULAS0034 and the $\sim$100~K giant planets of the Solar System.  Many of the
currently known extrasolar planets populate this temperature interval,
characterized by complex atmospheric physics: matter and radiation in these
cold, dense, and turbulent atmospheres couple into a very dynamical mix,
where molecules and dust form and dissipate. Current atmosphere models are
rather uncertain in this unexplored temperature range and they will
significantly benefit from observational constraints.
Two major physical transitions are expected to occur between $\sim$~700~K 
and $\sim$~400~K and strongly alter the emergent near-infrared 
spectra \citep{burrows.03}: NH$_3$ becomes an abundant 
atmospheric constituent and its near-infrared bands become
major spectral features, and water clouds form and deplete
H$_2$O from the gas phase. The corresponding near-infrared
spectral changes are likely to be sufficiently drastic
that the creation of a new spectral type will be warranted
\citep{kirkpatrick.2000}. \citet{kirkpatrick.1999,kirkpatrick.2000} 
reserved the ``Y'' letter for the name of that putative new spectral 
type.

To help fill this temperature gap, we conduct the Canada France 
Brown Dwarf Survey (CFBDS, \citep{delorme.2008}), which uses 
MegaCam \citep{Boulade.2003proc}  $i'$and  $z'$ images to select very 
cool brown dwarfs (and high redshift quasars) on their very red 
$i'-z'$ colour. We present here our coolest 
brown dwarf discovery to date, CFBDS J005910.90-011401.3 (hereafter 
CFBDS0059), a $>$T8 dwarf with evidence
for near-infrared NH$_3$ absorption. Section 2 describes 
its discovery, and presents our follow-up observations: $i'$, $z'$, $Y$, $J$, 
$H$ and $Ks$ photometry and astrometry  of CFBDS0059 and (as a reference) 
ULAS0034, laser guide star adaptive optics imaging 
and a near infrared spectrum of the new brown dwarf. 
Section 3 discusses the kinematics and the dynamical population
membership of CFBDS0059. Section~4 compares the 
spectrum of CFBDS0059 with those of Gl~570D (T7.5), 2MASS~J0415-09 
(T8) and ULAS0034, and in the light of synthetic spectra 
uses that comparison to determine its effective temperature, gravity
and metallicity. We also examine in that section the new spectral
features which appear below 700K, in particular an NH$_3$ band, 
and discuss new spectral indices for spectral classification beyond
T8. Finally, section 5 summarizes our findings and sketches 
our near-future plans.

\section{CFBDS}

Field ultracold brown dwarfs are intrinsically very faint, and 
as a result they can only be identified in sensitive wide-field 
imaging surveys. They are most easily 
detected in the near infrared, and one could thus 
naively expect them to be most easily identified in that wavelength 
range. Brown dwarf spectra however very much differ from a black body, and 
their considerable structure from deep absorption lines and bands 
produces broad-band pure near-infrared colours that loop back
to the blue. At modest signal to noise ratios, those colours
are not very distinctive. Brown dwarfs are therefore more easily 
recognized by including at least one photometric band under 1~$\mu$m. 
At those shorter wavelengths their spectra have extremely steep 
spectral slopes, and the resulting very red $i'-z'$ and $z'-J$ 
colours easily stand out. 

 As discussed in detail in \citet{delorme.2008}, the CFBDS survey 
brown dwarf identification is a two-step process:
\begin{itemize}
\item we first select brown dwarfs candidates on their red $i'-z'$ 
  colour in MegaCam images which cover several hundred square degrees.
\item $J$-band pointed observations of these candidates then discriminate 
  actual brown dwarfs from artefacts, and astrophysical 
  contaminants.
\end{itemize}

 The $i'-z'$ selection takes advantage of the  wide field of the
MegaCam camera 
\footnote{http://www-dapnia.cea.fr/Sap/Phys/Sap/Activites/Projets/Megacam}
on the 3.6m CFHT telescope, and of the  trove of deep observational 
material obtained with that instrument.  We rely on existing 
$i'$ images from the Very Wide component of the Canada France Hawaii 
Telescope Legacy Survey (CFHTLS-VW) and, for different fields,
on existing $z'$ images from the Red sequence Cluster Survey 2 (RCS2).
We then match those with either new $z'$ or new $i'$ exposures to 
obtain $i'$ and $z'$ pairs. The two parent surveys also obtain $g'$ and 
$r'$ images, which for the RCS2 survey are contemporaneous with the $z'$
exposure. We don't use those as primary selection tools, but 
the contemporaneous exposures from the RCS2 survey provide
a welcome check that an apparently red source was not, instead,
a variable brighter at all wavelenghts at the $z'$ epochs.
Since all our fields have low galactic extinction, the only other 
astrophysical point sources with a similarly red $i'-z'$ are quasars 
at redshifts of z$\approx6$, which represent the other motivation of 
our program \citep[e.g.][]{willott.2007}. We discriminate between 
quasars and  brown dwarfs with $J$-band photometry obtained on several 2 to 
4m-class telescopes \citep[for details]{delorme.2008}. These targeted 
follow-up observations also reject a number of remaining unflagged 
artefacts and they provide a refined spectral type estimate, thanks to 
the much higher signal-to-noise ratio which we typically achieve on 
$z'-J$ than on $i'-z'$.

\section{Observations}

We first identified CFBDS0059 as a brown dwarf candidate when comparing a 
360~s RCS2 $z'$ exposure from 2005 December 27 with a 500~seconds CFBDS
$i'$ exposure from 2006 August 31. CFBDS0059 is undetected in the 
$i'$~image to $i_{AB}'=25.2$ ($5\sigma$), in spite of a strong 
$z'$ detection ($z_{AB}'=21.93{\pm}0.05$).  The RCS2 survey obtains
contemporaneous $g'$, $r'$ and $z'$ images, and we checked the $g'$ 
and $r'$ exposures for a counterpart. These images, which 
were obtained within 50 minutes of the $z'$ observation,  
show no object at the position of CFBDS0059. This essentially 
excludes that the $z'$ detection was due to a variable or slowly 
moving object with neutral colours. The $i_{AB}'-z_{AB}'>3.2$  
($5\sigma$) lower limit was thus secure, and made CFBDS0059 a 
very strong candidate for follow-up.

\begin{figure}
\includegraphics[width=9cm,angle=0]{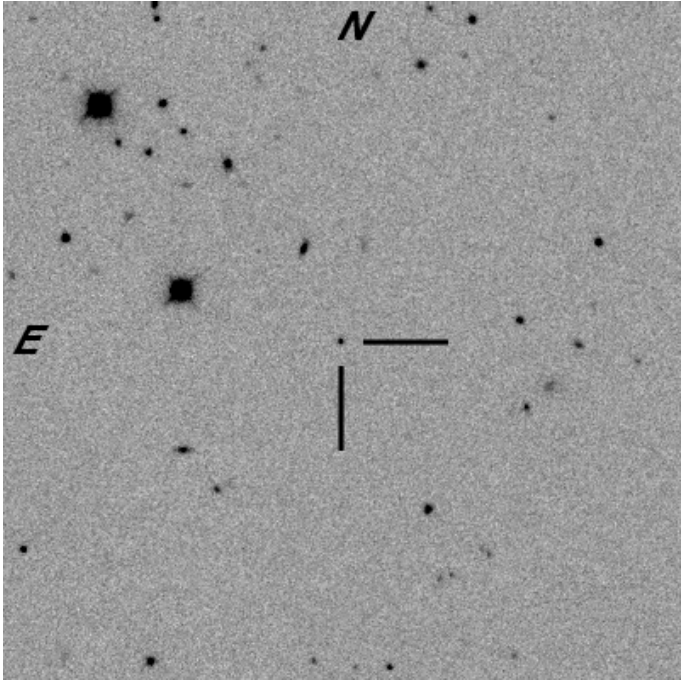}
\caption{$K_S$-band finder chart for CFBDS0059. The field of view 
is 2 by 2 arcminutes.}
\label{figure_izyjhk}
\end{figure}

\subsection{$J$-band photometry and near-infrared spectroscopy}

Our initial $J$-band imaging consists of five 20-second dithered 
exposures with the SOFI near-infrared camera on the ESO NTT 
telescope at La Silla on 2006 November 12. As discussed
below, the photometric system of that instrument is non-standard.
We used a modified 
version of the jitter utility within the ESO Eclipse package 
\citep{devillard.97} to subtract the background and coadd the five
exposures. We extracted photometry from the resulting image using 
PSF fitting within Source Extractor (\citet{Bertin.1996} and Bertin \& 
Delorme, in preparation) and obtain $J_{Vega}$=$18.11{\pm}0.06$. 
The resulting $z_{AB}-J_{AB}=3.0$ colour confirmed CFBDS0059 as a strong 
very late-T dwarf candidate and we triggered $H$-band spectroscopic
observations with NIRI \citep{hodapp.03} at Gemini-North. 
Those were obtained in queue mode on 2007 July 30 and immediately
confirmed the very cool nature of CFBDS0059. We then requested
$J$ and $K$-band spectroscopy with the same instrument, which
was obtained on 2007 September 1.

All spectra were obtained through a $\sim0\farcs7$-wide slit, which 
produces a resolving power of $\frac{\lambda}{\Delta\lambda}\sim500$. 
The object is dithered along the slit. The $H$-band spectrum is the sum
of 16 $300-$second integrations,   
while the $Y~+~J$ and $K$ band spectra each are the sum of 9 
$300-$second integrations. 
Consecutive image pairs are pair subtracted, flat fielded using a
median combined spectral flat and corrected for both spectral and
spatial distortions. Spectra are extracted using a positive and a
negative extraction box matched to the trace profile. A first wavelength
calibration was obtained with argon lamp arc spectra taken at the end of
the sequence, and the wavelength scale was then fine tuned to match the
atmospheric OH-lines. Individual spectra extracted from image pairs
were then normalized and median combined into final spectra. Per-pixel
S/N of 25, 40 and 5 where achieved on the J, H and K-band peaks
respectively.
For all 3 wavelength settings the A0 star HIP10512 was observed 
immediately after the science observations to calibrate the
instrumental spectral response and the telluric transmission.

\subsection{Additional near-infrared and optical Photometry}

\begin{table*}
\caption{Wide band $i'_{AB}$ and $z'_{AB}$ CFHT photometry  
measured with Megacam, MKO-system $Y$, $J$, $H$, and $K_s$ measured 
with WIRCam and associated colors.
For comparison we include the \citet{warren.2007},  $Y$, $J$, $H$, and
$K$ photometry of ULAS0034. 
}
\begin{tabular}{ccccccccc}
\hline
Target &  \multicolumn{7}{c}{Spectral band} \\
                & $i'_{AB}$ & $z'_{AB}$ & $Y$ & $J$ & $H$& $K_s$ &$K$ \\ \hline
                CFBDS0059  	& $>25.17{\pm}0.03$ 	& $21.93{\pm}0.05$  	& $18.82{\pm}0.02$ 	& $18.06{\pm}0.03$	& $18.27{\pm}0.05$	& $18.63{\pm}0.05$ & -\\
ULAS0034	& $>25.14{\pm}0.04$ 	& $22.11{\pm}0.06$  	& $19.12{\pm}0.03$ 	& $18.21{\pm}0.03$  	& $18.49{\pm}0.05$      & $18.33{\pm}0.05$ & -   \\ 
\citet{warren.2007} &                   &                	& $18.90{\pm}0.10$ 	& $18.15{\pm}0.03$  	& $18.49{\pm}0.04$      & - & $18.48{\pm}0.05$  \\ \hline
                &  $z'_{AB}-Y$    &  $Y-J$          &  $J-H$                &     $H-Ks$ &$H-K$ & & \\
CFBDS0059  	& $3.11{\pm}0.05$ & $0.76{\pm}0.04$ & $-0.21{\pm}0.06$ & $-0.36{\pm}0.07$ & - &   & \\
ULAS0034	& $2.99{\pm}0.07$ & $0.91{\pm}0.04$ & $-0.28{\pm}0.06$ & $0.16{\pm}0.07$  & - &   & \\
\citet{warren.2007} &             & $0.75{\pm}0.10$ & $-0.34{\pm}0.05$ & -  & $0.01{\pm}0.07$ &   & \\ \hline

\end{tabular}
\label{table_photo}
\end{table*}

The $J$ filter of the SOFI camera on the NTT has a quite non-standard 
bandpass, for which the large colour corrections that result from
the strong structrure in T dwarf spectra \citep[e.g.][]{stephens.04} 
have not been fully characterized. To compare the 
spectral energy distribution (SEDs) of CFBDS0059 and ULAS0034 
\citep{warren.2007} we therefore prefered to obtain additional 
near-infrared wide-band photometry with WIRCam \citep{puget.2004} 
on CFHT, which uses standard Mauna Kea Observatory infrared filters 
\citep[MKO system]{simons.02,tokunaga.02,tokunaga.05}. The observations 
(2007 August 1st and 5th) used 
dithering patterns of $\sim60$~arcsec amplitude for total (respectively
individual) exposure times of 300 (60), 150 (30), 300 (15) and 720 (20) seconds
for the $Y$, $J$, $H$ and $K_s$ bands. The skies were photometric and 
the seeing varied between 0.8 and 1.0''.

Table~\ref{table_photo} summarizes the magnitudes of CFBDS0059 and
ULAS0034 in all available bands. The WIRCam photometry
of ULAS0034 agrees with the \citet{warren.2007} measurements
within better than 1$\sigma$ for $H$ band and  within 1.5 $\sigma$
 for $J$.
 The \citet{warren.2007} K-band measurement used a
$K$ filter, while our WIRCam measurement uses the narrower
and bluer $Ks$ filter. The 0.15 mag difference between these two
observations is approximately consistent with the
\citet{stephens.04} prediction for the effect of these
different filter bandpasses at late-T spectral types.
Similarly, the better short-wavelength quantum efficiency
of the WIRCam detector can qualitatively explain the
0.2~magnitude (2$\sigma$) discrepancy between our
$Y$ photometry and the \citet{warren.2007} WFCam measurement.   
The near-IR colours of the two brown dwarfs are similar, except 
$H-Ks$ which is $\sim$0.5~magnitude bluer for CFBDS0059 than for
ULAS0034. We will interpret the implications of this low $Ks$ flux when
we examine the near-infrared spectrum.

\subsection{Astrometry}

CFBDS0059 and ULAS0034 are serendipitously just 6.3 degrees apart 
on the sky, and at similar photometric distances from Earth since 
they have similar spectral types and apparent magnitudes , and we 
initially entertained the idea that they might, perhaps, be part of 
a common moving group. 
The proper motion of CFBDS0059 however, measured between our 2005 
Megacam $z'$ and 2007 WIRCam $K_s$ images, and uncorrected for its
parallactic motion, is $\mu_{\alpha}=+0.94\pm0.06$"/yr, 
$\mu_{\delta}=+0.18\pm0.06$"/yr (Table~\ref{astrom}).
ULAS0034 on the other hand moves by $\mu_{\alpha}=-0.12\pm0.05$"/yr and
$\mu_{\delta}=0.35\pm0.005$"/yr \citep{warren.2007}. 
The two proper motions are thus sufficiently different that the two
brown dwarfs are clearly unrelated. We checked for main sequence 
common proper motion companions to CFBDS0059, which would have 
provided welcome age and metallicity constraints \citep[e.g.][]{scholz.2003}
but did not find any match within a 10 arcminute radius.

\begin{table}
\caption{Astrometry of CFBDS0059. (Epoch:August 5th, 2007)}
\begin{tabular}{llcc} \hline
RA(J2000) & DEC(J2000) & $\mu_{\alpha}$ & $\mu_{\delta}$ \\
hh:mm:ss & dd:mm:ss & \multicolumn{2}{c}{$``/yr$} \\ \hline
00:59:10.903 & -01:14:01.13 & $0.94\pm0.06$ &  $0.18\pm0.06$ \\ \hline
\end{tabular}
\label{astrom}
\end{table}

\subsection{Keck Laser Guide Star Adaptive Optics Imaging}
To search for binarity, we imaged CFBDS0059 on 16~January~2008~UT
using the laser guide star adaptive optics (LGS AO) system
\citep{vandam.2006,wizinowich.2006} of the 10-meter Keck
II Telescope on Mauna Kea, Hawaii.  Conditions were photometric with
better than average seeing. We used the facility IR camera NIRC2 with its
narrow field-of-view camera, which produces an image scale of
$9.963\pm0.011$~mas/pixel \citep{pravdo.2006} and a
$10.2\arcsec \times 10.2\arcsec$ field of view.  The LGS provided the
wavefront reference source for AO correction, with the exception of
tip-tilt motion.  The LGS brightness was equivalent to a
$V\approx9.8$~mag star, as measured by the flux incident on the AO
wavefront sensor.  Tip-tilt aberrations and quasi-static changes in
the image of the LGS as seen by the wavefront sensor were measured
contemporaneously with a second, lower-bandwidth wavefront sensor
monitoring the $R=14.6$~mag field star USNO-B1.0~0887-0010532
\citep{monet.2003}, located 32\arcsec\ away from CFBDS0059.
The sodium laser beam was pointed at the center of the NIRC2
field-of-view for all observations.

We obtained a series of dithered images, offsetting the telescope by a
few arcseconds, with a total integration time of 1080s.  We
used the $CH4s$ filter, which has a central wavelength of
1.592~$\mu$m and a width of 0.126~$\mu$m.  This filter is
positioned near the $H$-band flux peak emitted by late-T~dwarfs.
The images were reduced in a standard fashion.  We constructed flat
fields from the differences of images of the telescope dome interior
with and without continuum lamp illumination.  Then we created a
master sky frame from the median average of the bias-subtracted,
flat-fielded images and subtracted it from the individual images.
Images were registered and stacked to form a final mosaic, with a
full-width at half-maximum of 0.09\arcsec\ and a Strehl ratio of 0.05.
No companions were clearly detected in a $6\arcsec \times 6\arcsec$
region centered on CFBDS0059.

We determined upper limits from the direct imaging by first smoothing
the final mosaic with an analytical representation of the PSF's radial
profile, modeled as the sum of multiple gaussians.  We then measured
the standard deviation in concentric annuli centered on the science
target, normalized by the peak flux of the targets, and adopted
10$\sigma$ as the flux ratio limits for any companions.  These limits
were verified with implantation of fake companions into the image
using translated and scaled versions of the science target.  

\begin{figure}
\includegraphics[width=7cm,angle=90]{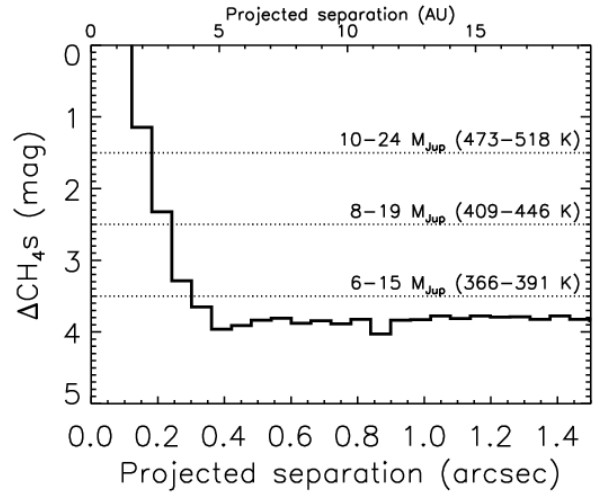}
\caption{Adaptive Optic detections limits as a function of the separation
  from CFBDS0059. Dashed lines show the equivalent mass and
  temperature for a given contrast. The first value assumes
  an age of 1 Gyr and the second an age of 5 Gyr.
}
\label{AOlimits}
\end{figure}

Figure \ref{AOlimits} presents the final upper limits on any companions.  We
employed the ''COND'' models of \cite{baraffe.03} to convert
the limits into companion masses, for assumed ages of 1~and 5~Gyr and
a photometric distance estimate of 13~pc.  We assumed any cooler
companions would have similar $(CH4s-H)$ colors to
CFBDS0059.

\section{Kinematics}

\begin{figure}
\includegraphics[width=9cm,angle=0]{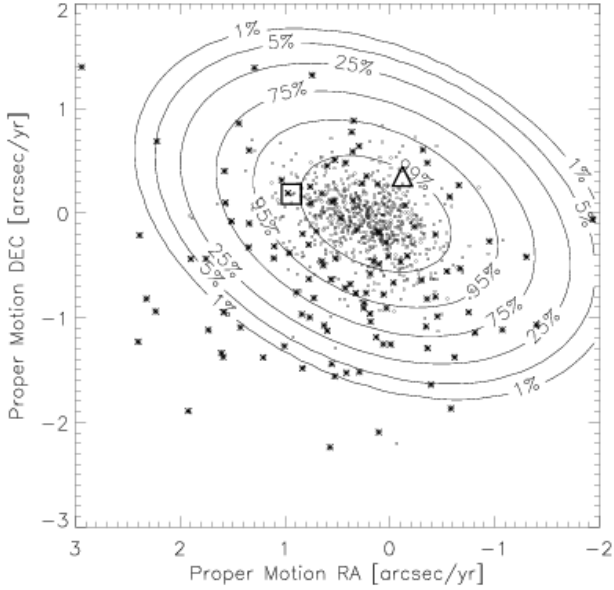}
\caption{Thin disk probability membership contours in proper motion
space from the Besancon stellar population model. The contours are
generated for synthetic stars with distances between 10 and 20 pc, 
belonging to the thin disk (small dots) and the thick disk (small 
stars, with the density of the latter increased by a factor of 10 
for display purposes). 
Less than one halo star would appear on the  plot. Based on their
measured proper motion, the likelihood that CFBDS0059 (large open 
square) and ULAS0034 (large open triangle) belongs to the thin disk are
95\% and $>$ 99\%.}
\label{figure_pm}
\end{figure}

We estimate a spectrophotometric distance for CFBDS0059 by adopting
M$_J=17.5\pm0.5$, based on an approximate T9/Y0 spectral type (discussed below)
and on an extrapolation of the M$_J$ versus spectral type relation of 
\citet{knapp.04} beyond T8 (2MASSJ$0415-0935$, hereafter 2M0415). 
This extrapolation is consistent with the \citet{chabrier.00} models 
which predict $\Delta J \sim 1.0$ between brown dwarfs of  
$T_\mathrm{eff} \sim$ 750K and  625K (like 2M0415 and CFBDS0059).
The resulting d~=~$13\pm5$~pc has significant systematic uncertainties,
because spectral typing beyond T8 is just being defined, and especially 
because the linear 1~magnitude/subtype decline seen at earlier
subtypes may not continue beyond T8.  The adaptive optics 
observations exclude any companion of similar luminosity beyond 1.2 AU, 
but CFBDS0059 could still of course be a tighter binary. Its small 
distance fortunately puts CFBDS0059 within easy reach of
modern parallax measurements. 

We use the Besancon stellar population model \citep{robin.2003} to 
generate synthetic stars between 10~pc and 20~pc for the thin
(dots) and thick (star symbol) disk populations at the galactic
position of CFBDS0059. Fig~\ref{figure_pm} shows their
proper motions together with those of CFBDS0059 and
ULAS0034. The contour lines show the probability that an object
with a given proper motion belongs to the thin disk rather than the 
thick disk (halo membership probabilities are negligible). CFBDS0059,
at its probable distance,  is well within the 95\% probability thin 
disk membership region, and ULAS0034 is within the 99\% probability 
region. In spite of its somewhat high proper motion for an object
beyond 10~pc, CFBDS0059 therefore most likely belongs to the thin disk.
The mean age of the simulated stars in the region of the
proper motion diagram occupied by CFBDS0059 is 4 Gyr,
suggesting that it is an older member of the thin disk. 
That age is consistent with the 1 to 5 Gyr range 
derived below from comparison to COND evolutionary
models \citep{baraffe.03}.  As any kinematic age for an 
individual star, this determination has large error bars, but 
it suggests that CFBDS0059 might be older than ULAS0034.

\section{Spectral comparison and atmospheric parameters}

\begin{figure*}
\includegraphics[width=13cm,angle=270]{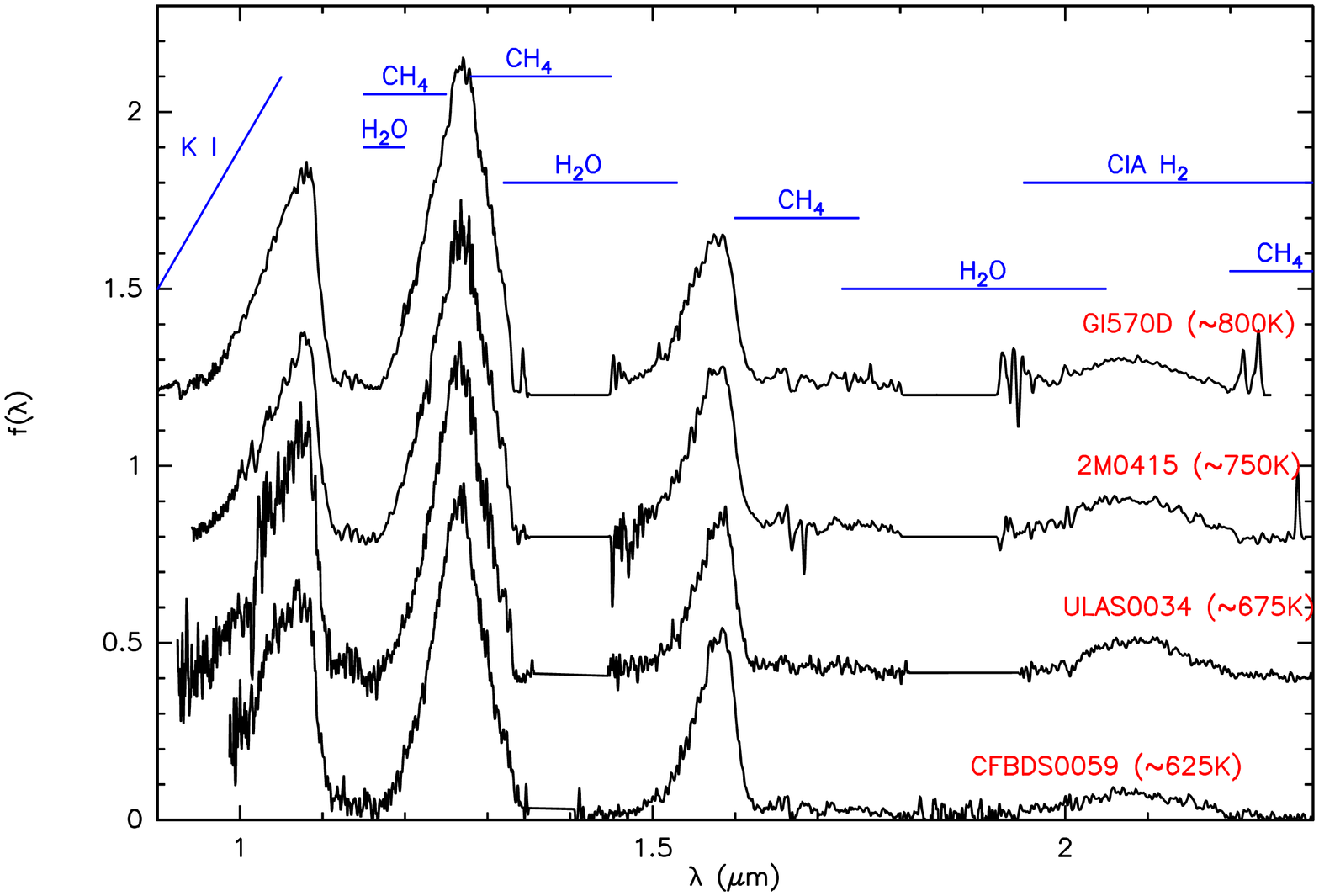}
\caption{0.9 $\mu$m - 2.3 $\mu$m spectra of CFBDS0059 and the
three other coolest brown dwarfs. The spectra are normalized 
to unit flux densities at their 1.27$\mu$m.peak, and vertically offset 
for clarity. 
The main T-dwarf spectral features are labeled. The 
temperatures of Gl~570D and 2M0415 are from the
careful spectroscopic analyses of \citet{saumon.06} and 
\citet{saumon.07}. Those of CFBDS0059 and ULAS0034 
are from our $W_J$ versus $J/K$ index (Fig.~\ref{wj_jk}). 
}
\label{spe_compa}
\end{figure*}

\begin{figure*}
\begin{tabular}{ccc}
\includegraphics[width=6cm,angle=0]{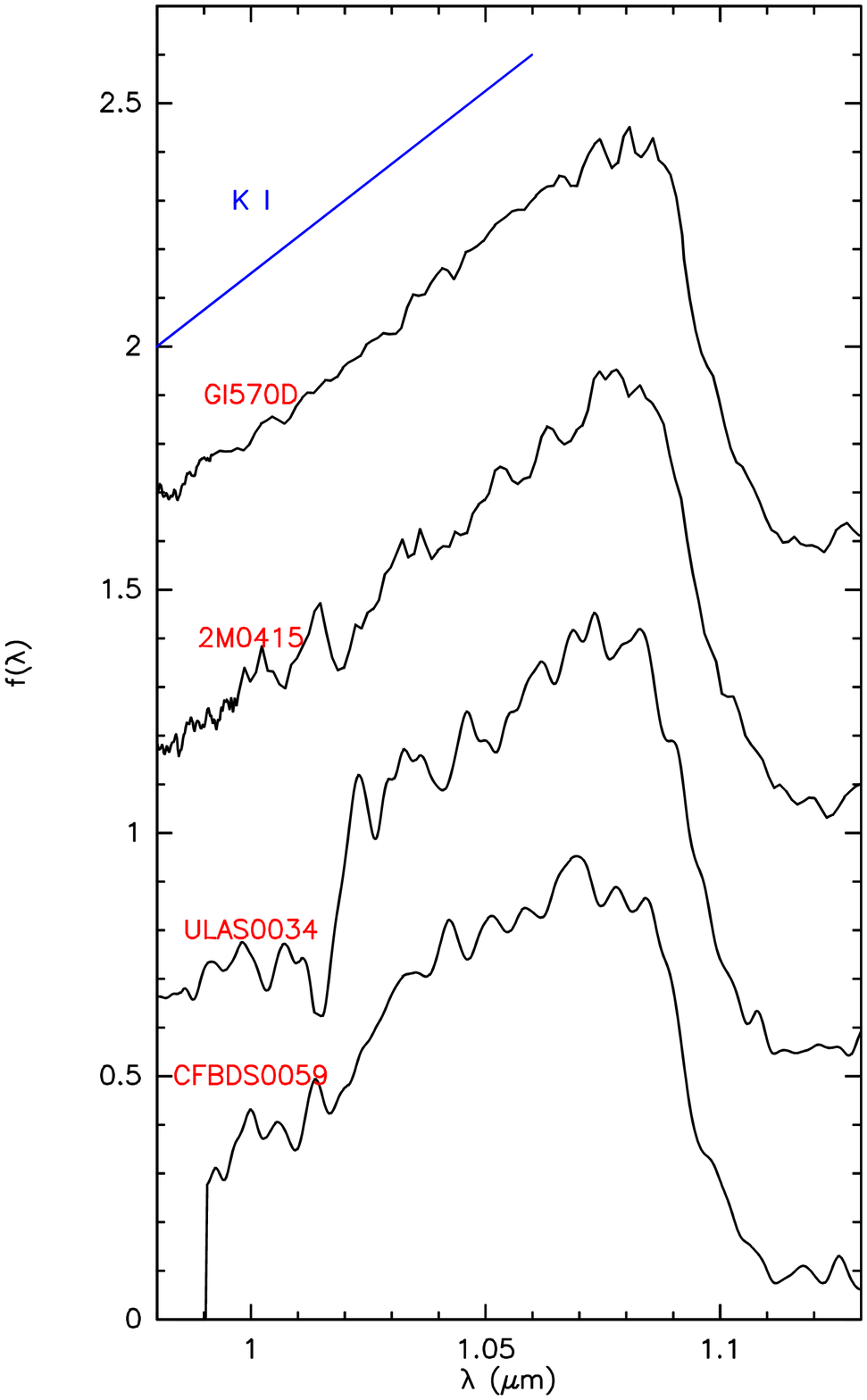} &
\includegraphics[width=6cm,angle=0]{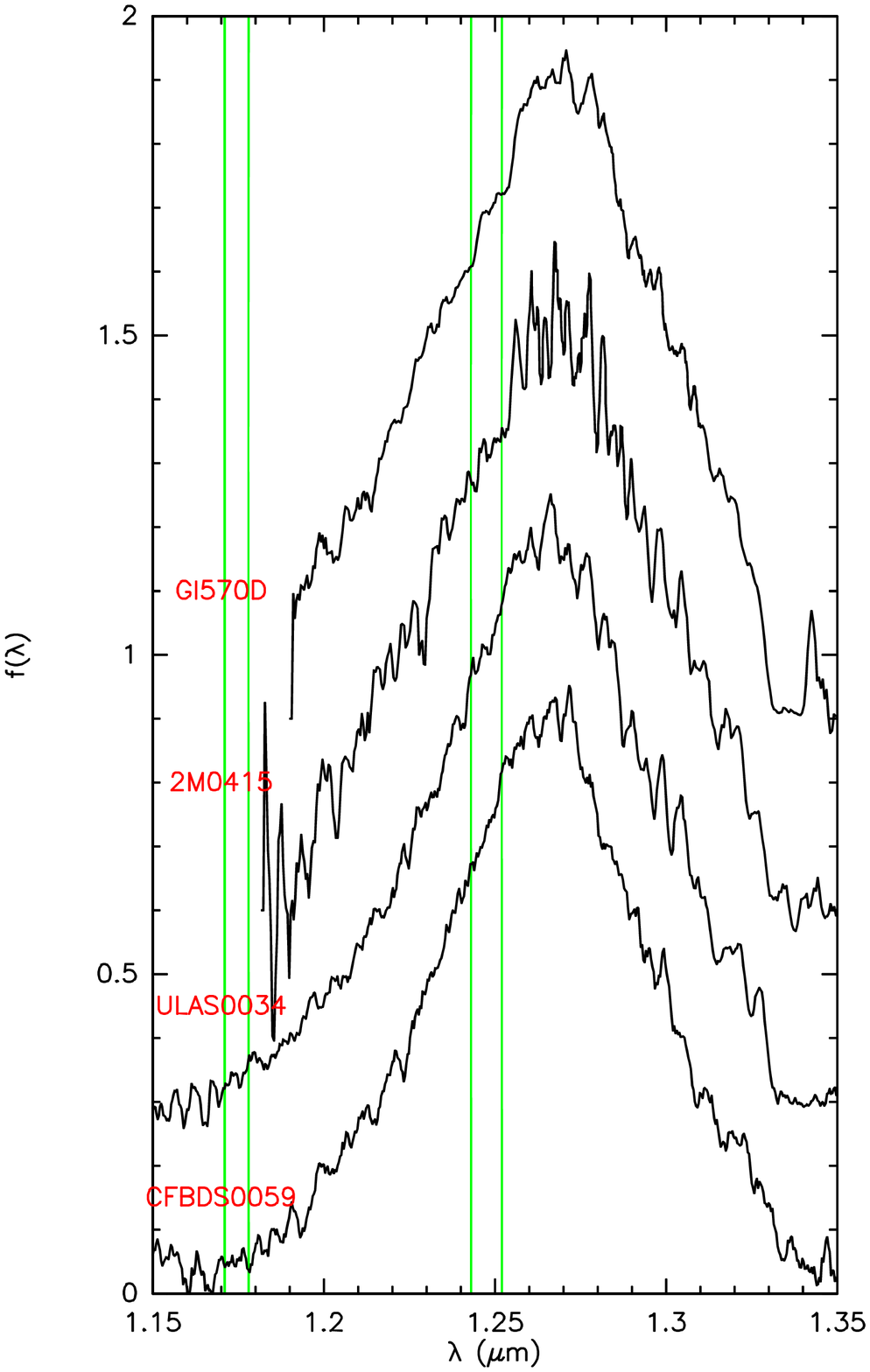} &
\includegraphics[width=6cm,angle=0]{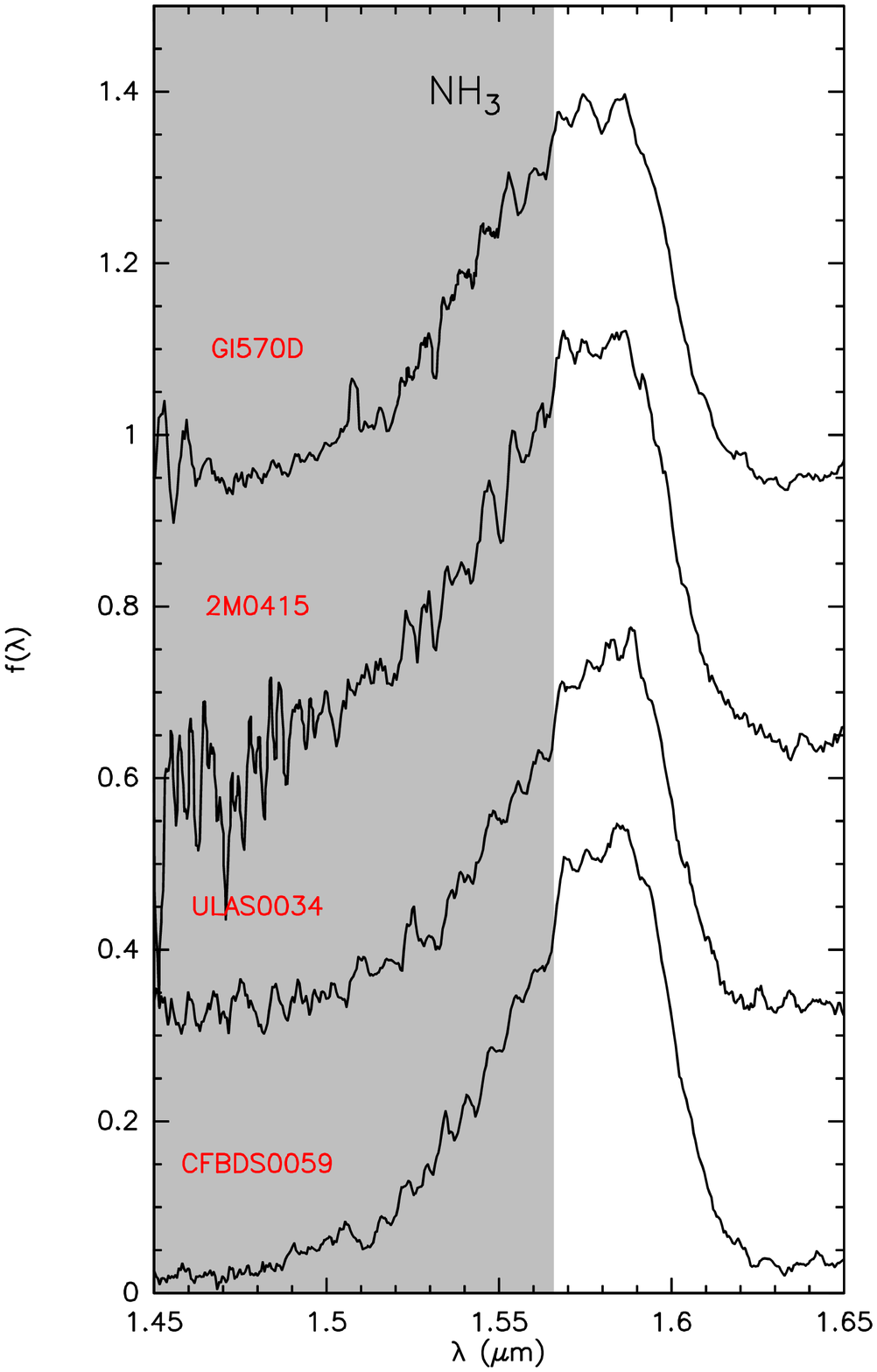} 
\end{tabular}
\caption{$Y$, $J$ and $H$-bands 
spectra of the four cool brown dwarfs. The green vertical lines in the 
1.15 $\mu$m - 1.35 $\mu$m panel mark the two potassium doublets.
The grey band in the 1.45 $\mu$m - 1.65 $\mu$m panel indicates
the approximate limits of the new absorption bands, which we discuss 
in the text.}
\label{spe_compa_band}
\end{figure*}

Fig~\ref{spe_compa} and Fig~\ref{spe_compa_band} present our 
spectrum of CFBDS0059, together with those of ULAS0034 
\citep[$>$T8]{warren.2007}, 2M0415 
\citep[T8]{burgasser.03} and Gl~570D  \citep[T7.5]{burgasser.2000}, 
which successively were the coolest known brown dwarfs. 
Thanks to their earlier discovery, 2M0415 and 
Gl~570 have the best characterized atmospheric parameters
\citep{saumon.06,saumon.07}, and they provide the most solid baseline
for a differential study. The \citet{warren.2007} spectrum of 
ULAS0034, kindly communicated by S.~Leggett, was obtained
with GNIRS on Gemini South and its $\frac{\lambda}{\Delta\lambda}$~=~500
resolution matches that of our NIRI spectrum of CFBDS0059.
We downloaded the \citet{burgasser.03,burgasser.02} OSIRIS spectra 
of the two other brown dwarfs from the Ultracool Dwarf 
Catalog\footnote{http://www.iac.es/galeria/ege/catalogo\_espectral/index.html},
and degraded their original spectral resolution of 
$\frac{\lambda}{\Delta\lambda}{\sim}1200$ to match that of the
GNIRS and NIRI spectra. Stronger telluric absorption from 
the lower altitude telescopes explains the wider blanked 
regions in the corresponding spectra, but doesn't measurably 
affect any comparison: as illustrated by CFBDS0059, late-T dwarfs 
have essentially negligible flux wherever telluric H$_2$O absorption 
matters.  Because the OSIRIS spectra do not cover the $Y$ 
band, we complement them by lower resolution spectra from 
\citet{geballe.01} and \citet{knapp.04} for $\lambda < 1.18 \mu m$.

\subsection{Atmospheric parameters}

Atmospheric parameters of ultracool dwarfs are ideally determined
from a combination of near and mid-IR information 
\citep[e.g.][]{saumon.06,saumon.07}, but low resolution near-infrared 
spectra alone provide a useful proxy when mid-IR photometry and
spectra are not (yet) available \citep[e.g.][]{burgasser.2006,leggett.2007}.
\citet{burgasser.2006} used a grid of solar metallicity cool brown dwarfs 
to calibrate two spectral ratios,  H$_2$O-$J$ and $K/H$, which respectively 
measure the strength of H$_2$O absorption at $\sim$ 1.15 $\mu$m and
the flux ratio between the $K$ and $H$ peaks, to T$_{\rm eff}$ and
$\log g$. \citet{warren.2007} however found that H$_2$O-$J$ essentially 
saturates below T$_{\rm eff}$~=~750K, and therefore chose not to use this
spectral index for spectral types later than T8.
They demonstrate on the other hand that the combination
of the $K/J$ index with the width of $J$-band peak, parametrised by 
their $W_J$ index, becomes a good T$_{\rm eff}$ and $\log g$ diagnostic
at  T$_{\rm eff}\simeq$900~K, and remains useful significantly 
below ~750K. We adopt their method.

\begin{figure}
\includegraphics[width=8.9cm,angle=0]{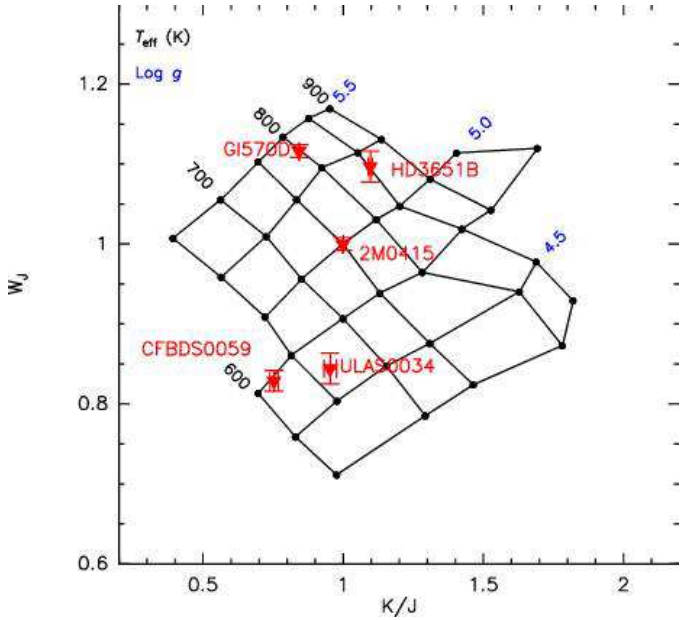}
\caption{$W_J$ versus $J/K$ indices. The grid represents 
indices measured on solar metallicity BT-settl model 
spectra, shifted into agreement of the [T$_{\rm eff}$=750K; 
$\log g$=5.00 and [M/H]=0] with the measured indices of
2M0415. The error bars represent the 1$\sigma$ uncertainties 
of the measured spectral indices.}
\label{wj_jk}
\end{figure}

Table~\ref{table_indice} lists our measurement of these two indices
for CFBDS0059, and Fig~\ref{wj_jk} compares them with the 
\citet{warren.2007} measurements for Gl~570D, HD3651B, 2M0415 and
ULAS0034. To derive T$_{\rm eff}$ and $\log g$ from the indices
we use model indices from solar-abundance BT-settl atmospheric
models \citep[Allard et al. 2008 in prep]{warren.2007,allard.03}. 
 The model with NH$_3$ at chemical equilibrium abundance 
clearly produces too much absorption in the blue side of the $H$-band,
confirming the finding of \citet{saumon.06,saumon.07}  that
non-equilibrium processes keep  the NH$_3$ partial pressure well below
its equilibrium  value. We then use models that keep the abundances of
NH$_3$ and N$_2$ at a fixed value in all parts of the atmosphere where the
reaction timescale exceeds the mixing 
timescale, which typically occurs at the 600-800K temperature level. 
These "quenched" models agree much better with the observed band shape.

As a first 
order correction for the remaining imperfections
of the theoretical spectra, the model indices are shifted into
agreement with the measurements of 2M0415 at the 
[T$_{\rm eff}$=750K; $\log g$=5.00 and [M/H]=0] determined for
that brown dwarf by \citet{saumon.07}. The T$_{\rm eff}$=800K and
$\log g$=5.35 resulting from this calibration for Gl~570D are
consistent with the T$_{\rm eff}$=800-820K and $\log g$=5.1-5.25 
derived by \citet{saumon.06} from a complete spectral analysis.
For HD3651B, T$_{\rm eff}$=820-890K and
$\sim \log g$=5.1-5.3 resulting from this calibration are roughly
consistent with the T$_{\rm eff}$=780-840K and $\log g$=5.1-5.5 
derived by \citet{liu.07}.

CFBDS0059 and ULAS0034 have very similar $W_J$ indices,
but the new brown dwarf has a significatively smaller $K/J$ index. 
Visual comparison of the two spectra (Fig.~\ref{spe_compa}) 
confirms that CFBDS0059 does have a weaker $K$-band 
peak than any of the 3 comparison cool brown dwarfs. As widely
discussed in the recent literature (e.g. \citet{liu.07},  Fig.~3 in  
\citet{burgasser.2006}, or Fig.~3 in \citet{leggett.2007}),
for a fixed metallicity a weaker $K$-band peak is evidence for
either a lower temperature or a higher gravity. The $W_J$
index lifts this degeneracy: it indicates, again assuming 
identical chemical compositions for the two brown dwarfs, 
that CFBDS0059 is cooler by $\sim$50$\pm$15~K and has a 
$\sim$0.15$\pm$0.1 higher $\log g$  than ULAS0034.

As also discussed by \citet{warren.2007}, the above uncertainties 
only reflect the random errors in the spectral indices. They are 
appropriate when comparing two very similar objects, like CFBDS0059
and ULAS0034, since systematic errors then cancel out. They must 
on the other hand be increased to compute absolute effective 
temperatures and gravity: one then needs to account for the 
uncertainties on the 2M0415 parameters which anchor the 
Fig~\ref{wj_jk} grid (T$\sim \pm 25$K and $\log g \sim \pm 0.2$; 
\citet{saumon.07}), and for the uncertainties in the atmospheric
models which may distort the grid between its anchor point 
[T$_{\rm eff}$=750K; $\log g$=5.00 and [M/H]=0] and the $\sim$600K
region of interest here. We conservatively adopt 
T$_{\rm eff}$=620$\pm$50K and $\log g$=4.75$\pm$0.3.

This 2-parameter analysis obviously cannot determine all three 
main atmospheric parameters (T$_{\rm eff}$,, $\log g$ and metallicity).
As discussed by \citet{warren.2007}, it actually determines the temperature 
with no ambiguity but leaves a combination of $[M/H]$ and  $\log g$ 
undetermined, and they demonstrated that in the $W_J$ versus $J/K$ plot 
metallicity is degenerate with surface gravity, with
$\Delta(\log g) \equiv -2\Delta[M/H]$. CFBDS0059 is thus definitely 
cooler than ULAS0034, but from  $W_J$ versus $J/K$ diagram it
can have either higher surface gravity {\it or} lower metallicity. 
This degeneracy affects the full JHK-band spectrum, where any metallicity
vs gravity difference is at most very subtle. It is however lifted by the 
shape of the $Y$-band peak (Figs.~3 of \citet{burgasser.2006} or 
\citet{leggett.2007}), since lower metallicity shifts the $Y$-band flux
density peak of submetallic brown dwarfs significantly blueward. 
Fig~\ref{spe_compa_band} shows no such shift, and the two objects
therefore have similar metallicities.

\begin{figure*}
\includegraphics[width=13.5cm,angle=270]{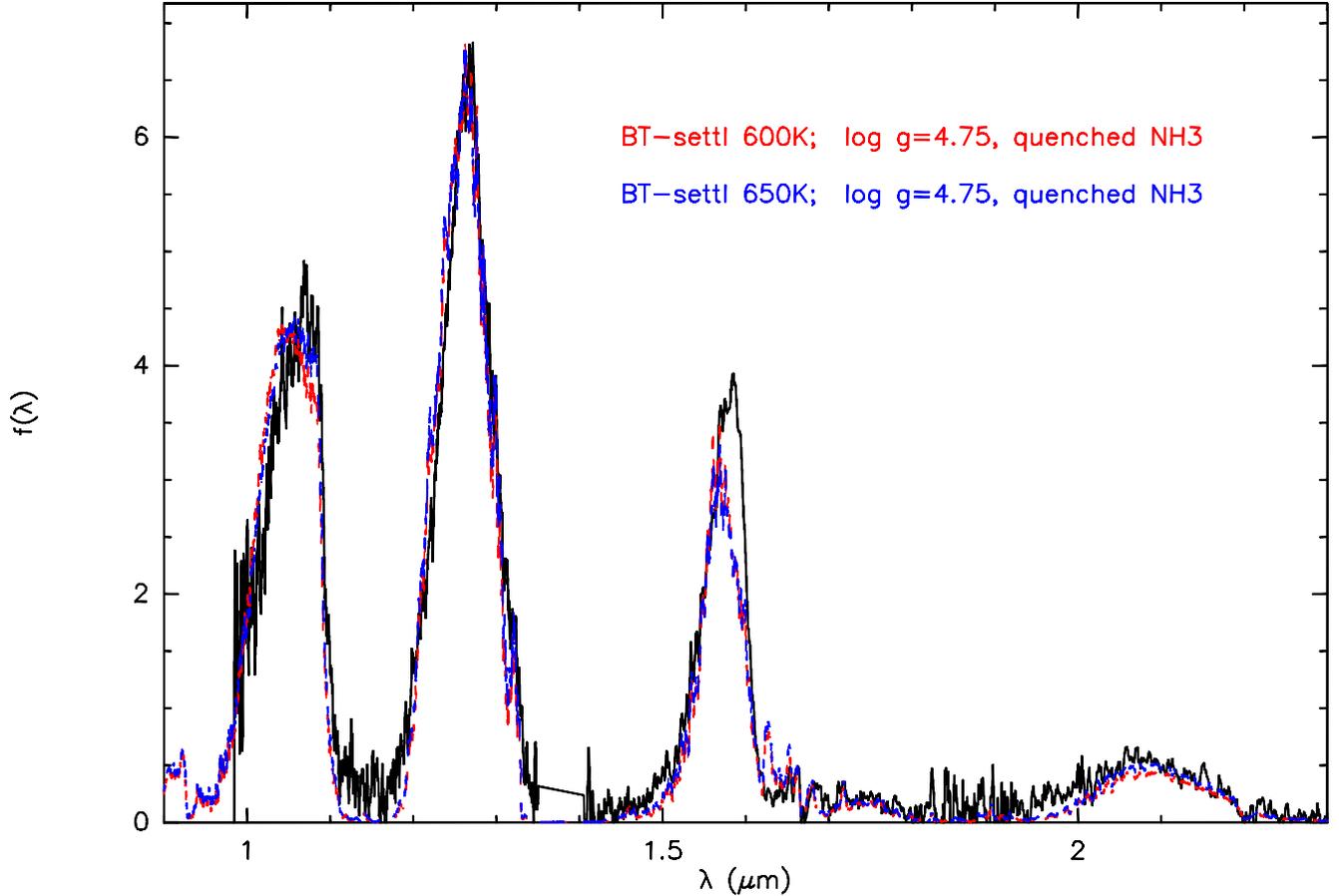}	
\caption{Overlay of the CFBDS0059 spectrum with the solar metallicity 
[T$_{\rm eff}$=600-650K; $\log g$=4.75] BT-settl 
synthetic spectrum. The two spectra are scaled to agree at their
1.27$\mu$m flux peak. The ``quenched NH3'' models are chemical
equilibrium models which enforce a constant abundance of ammonia in
the cooler regions of the atmosphere.  }
\label{spectre_fits}
\end{figure*}

Fig.~\ref{spectre_fits} overlays the observed CFBDS0059 spectrum with the 
synthetic spectrum for the closest point of the solar metallicity 
atmospheric model grid. Except on the red side of the $H$-band, 
model and observations agree well, boosting our confidence
in the derived atmospheric parameters. 
The main remaining predictive shortcoming of the models is their overestimated
absorption on the red side of the $H$-band peak. The principal 
opacity source in this region is the methane band centred at 
$1.67~{\mu}m$, for which comprehensive theoretical predictions 
are available, but only for transitions from the vibrational 
ground state (as will be discussed in detail in Homeier et al., in 
preparation). To make up for the missing absorption from higher 
bands, which constitutes a significant fraction of the opacity at brown 
dwarf temperatures, a constant empirical correction factor
was used. This correction must in turn lead to some overestimate 
of the CH$_4$ absorption as we reach the lower end of the T dwarf 
temperature range. Another possible source of errors are uncertainties 
in the models' temperature profile. The BT-Settl models 
self-consistently describe gravitational settling, convective turbulence,
advection, condensation, coalescence and coagulation of condensates 
to predict the formation and vertical extent of cloud layers 
\citep{allard.03,helling.08}. In late T dwarfs these clouds are 
predicted to reside deep in the optically thick part
of the atmosphere. Their opacity is thus not directly visible 
in the spectrum, but it may still
impact the thermal structure, and thus the relative abundance especially of
temperature-sensitive species like CH$_4$.

 Another (less serious) disagreement between the models and the observed
spectra occurs in the $Y$ band. The models overestimate the flux on 
the blue side of the $Y$-peak, and they imperfectly reproduce the 
general shape of the peak. As discussed below, the opacities in that
band are dominated by pressure-broadened wings of the 0.77~${\mu}m$ 
K~I line on the blue side and CH$_4$ on the red side.

For \citet{baraffe.03} evolutionary models, the
T$_{\rm eff}\simeq$570-670K and $\log g \simeq$ 4.45-5.05  
determined above translate into an age of 1-5~Gyr and a 
mass of 15M$_{Jup}$ (for 1~Gyr) to 30M$_{Jup}$ (for 5~Gyr).
The kinematics of CFBDS0059 suggests that it belongs to 
an older population, and therefore slightly favour a higher
mass and older age.

\subsection{Individual spectral features}
Fig.~\ref{spe_compa_band} zooms on the $Y$, $J$ and $H$-band
peaks of the four cool brown dwarf spectra. The published 
OSIRIS spectra of Gl~570D and 2M0415 do not cover the $Y$ band, 
which instead is plotted from the lower resolution spectra 
of \citet{geballe.01} and \citet{knapp.04}. For easier comparison,
the CFBDS0059 and ULAS0034 $Y$-band spectra are smoothed to that
resolution.

Direct comparison of the four spectra can be 
used to shed light on incipient new features and atmospheric chemistry. 
Features which are seen in both CFBDS0059 and ULAS0034 
are likely to be real even when their significance is modest in each
object, and those which are absent or weaker in the two hotter brown dwarfs,
can reasonably be assigned to low temperature molecules. Conversely,
features which disapear in the two cooler objects trace
higher temperature species.

As discussed above, the $Y$-band spectra of CFBDS0059 and ULAS0034 
do not differ much. Given the strong sensitivity of that band to
[M/H] that implies that the two objects have similar chemical 
compositions. The shape of the $Y$-band peaks of these two coolest 
brown dwarfs on the other hand differ from that of Gl~570D and 2M0415,
with the CFBDS0059 and ULAS0034 peaks extending further into the blue.
The dominant absorbers in the blue wings of the $Y$-band peak is 
the pressure-broadened wing of the 0.77 $\mu$m K~I line \citep[e.g.]
[]{burgasser.2006}, which must weaken as K~I depletes from the gas
phase under T$_\mathrm{eff}$=$\sim$700K. As anticipated by 
\citet{leggett.2007}, the slope of the blue side of the $Y$-band 
peak therefore shows good potential as an effective temperature 
diagnostics beyond spectral type T8.

The strength of the J-band K~I doublet is a good gravity estimator in
ultracool dwarfs \citep[e.g.][]{knapp.04}, because an increased
pressure at a fixed temperature favors KCl over K (\citet{lodders.99})
and consequently weakens atomic potassium features. At 
T$_{\rm eff} \sim$ 750-800K the the $J$-band K I doublets remain 
weakly visible and useful as a gravity proxy (Fig.~7 of \citet{knapp.04}). 
At T$_{\rm eff} < $ 700K on the other hand, the K~I doublets have 
completely vanished at the resolution of the current spectra
(Fig.~\ref{spe_compa_band}), even at the probably lower gravity of 
ULAS0034. Potassium is thus  mostly converted to KCl (or perhaps
other compounds) in the relevant photospheric layers.

\begin{figure*}
\begin{tabular}{cc}
\includegraphics[width=9.5cm,angle=0]{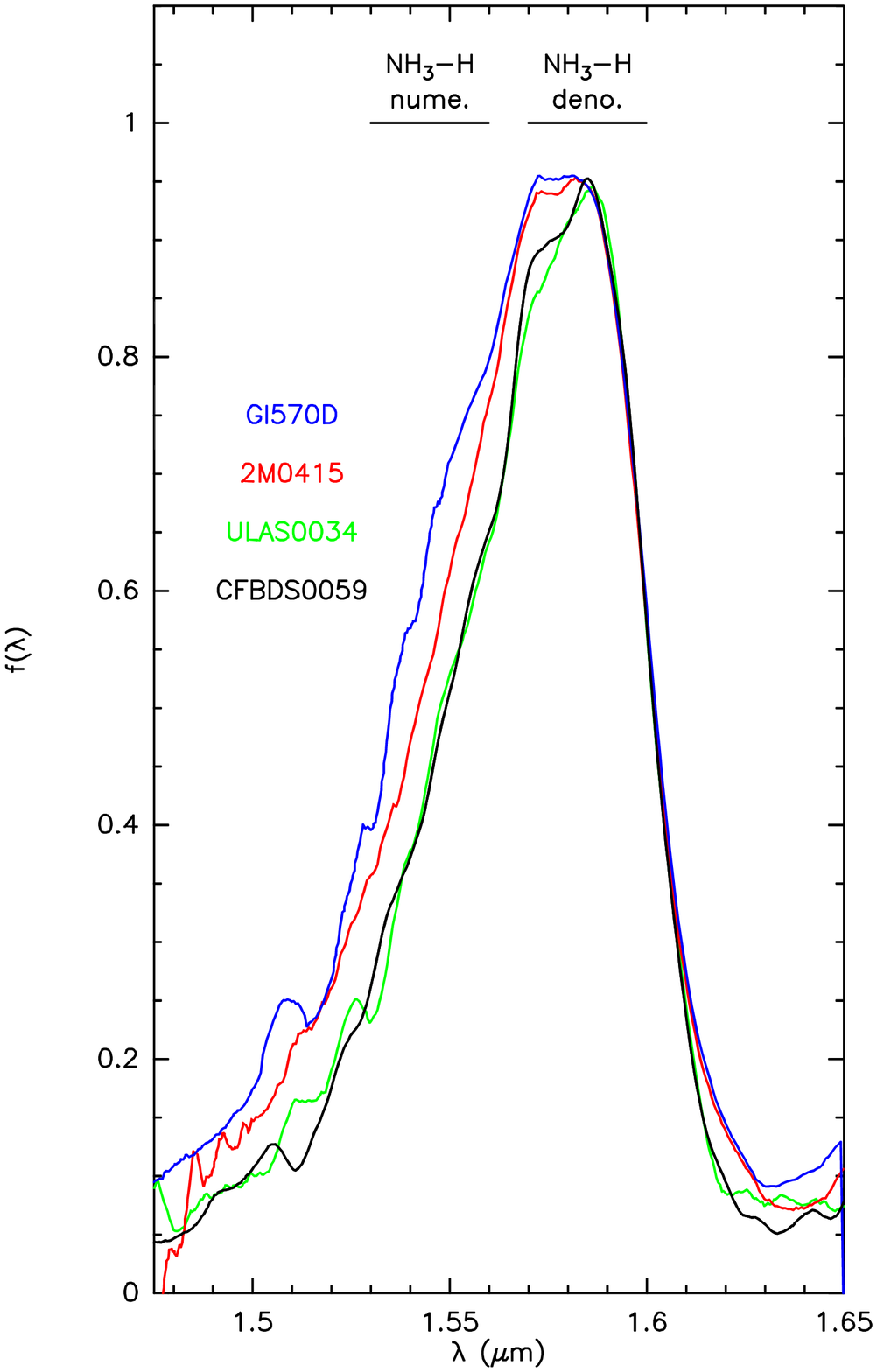} &
\includegraphics[width=9.5cm,angle=0]{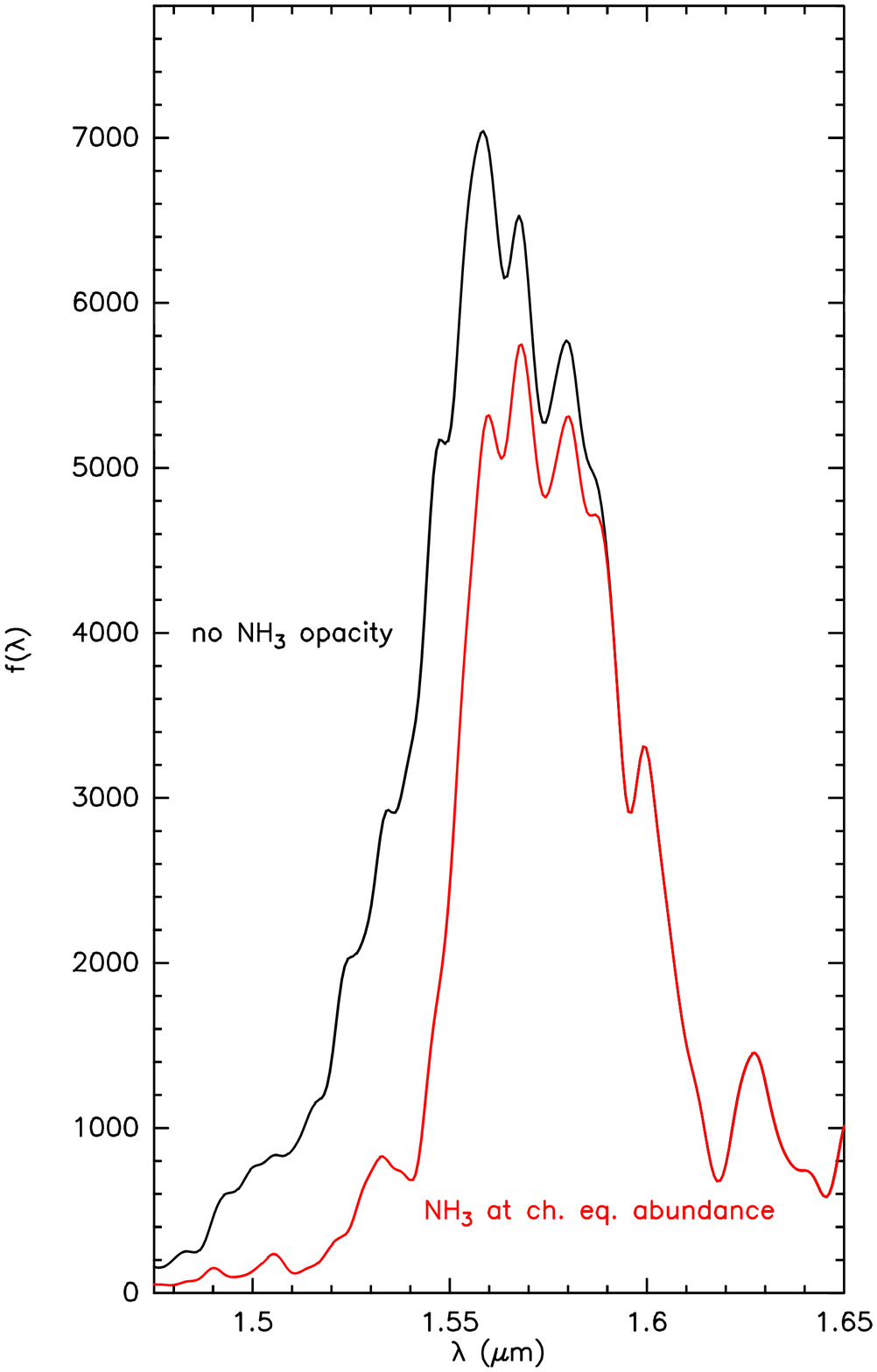}
\end{tabular}
\caption{{\bf Left}: $H$-band spectrum of the four cool brown dwarfs binned 
to R=$\sim$100. The spectra are normalized at $\lambda~=~1.59 \mu$m.
The integration intervals for the numerator and denominator of our 
proposed NH$_3-H$ index are marked.
{\bf Right}: BT-Settl synthetic spectra for 
[$T_\mathrm{eff} = 600$K; $\log g = 4.75$] with and without near-infrared
NH$_3$ opacity; the NH$_3$ abundance is at  its chemical equilibrium 
value.}
\label{spectre_comparaison_h2}
\end{figure*}

The strongest new feature is wide absorption on the blue side of the 
$H$-band,  at $\lambda < 1.565 \mu m$. It is conspicuous 
in CFBDS0059 and well detected in ULAS0034, and with hindsight 
is weakly visible in the 2M0415 spectrum 
(Fig.~\ref{spe_compa_band}). It is however clearly stronger 
at T$_{\rm eff} < $ 700K.  To visually emphasize this broad 
feature, we bin the spectra to R=$\sim$100 and overlay the four 
H-band spectra (Fig.~\ref{spectre_comparaison_h2}, left panel). Absorption 
sets in at $\sim 1.585 \mu m$ and becomes deeper for  
$\lambda < 1.565 \mu m$. These wavelengths overlap
with strong H$_2$O and NH$_3$ bands. Either molecule 
could a priori be responsible for the absorption.

\subsection{Near infrared ammonia signatures}

Those molecules unfortunately
have imperfect opacity data, and the NH$_3$ laboratory line lists in
particular are incomplete below  $1.7 \mu$m. Computed ammonia
opacities are therefore strictly lower limits. \citet{leggett.2007} 
compare synthetic spectra computed with and 
without NH$_3$ opacity, using the \citet{irwin.1999} line list for
$\lambda < 1.9\mu$m, and find that ammonia absorption in cold brown
dwarfs strongly depletes the blue wing of the $H$ band (their
Fig.10). Similarly, Fig.~9 of \citet{saumon.2000} plots synthetic $H$-band 
spectra with and without NH$_3$ opacity, and find  differences in two 
wavelength ranges: the NH$_3$-rich model is significantly more absorbed 
for $\lambda < 1.565 \mu$m and it has weaker but significant absorption
in the $[1.5725-1.585 \mu m]$ range. 

Fig~\ref{spectre_comparaison_h2} right panel plots two BT-Settl 
models for [$T_\mathrm{eff} = 600$K; $\log g = 4.75$], without any 
near-infrared NH$_3$ opacity, and with NH$_3$ opacity for that 
molecule at its chemical equilibrium abundance. As discussed
above the BT-Settl models do not reproduce the observed 
$H$-peak shape very well, and a quantitative comparison is thus
difficult. The comparison of the two models nonetheless confirms 
the \citet{saumon.2000} conclusion that ammonia produces strong
absorption below $\sim1.57\mu$m and weaker residual out to 
1.595$\mu$m. These model predictions qualitatively match the behaviour seen 
in Fig~\ref{spectre_comparaison_h2}, left panel.

\begin{figure}
\begin{tabular}{c}
\includegraphics[width=6.6cm,angle=270]{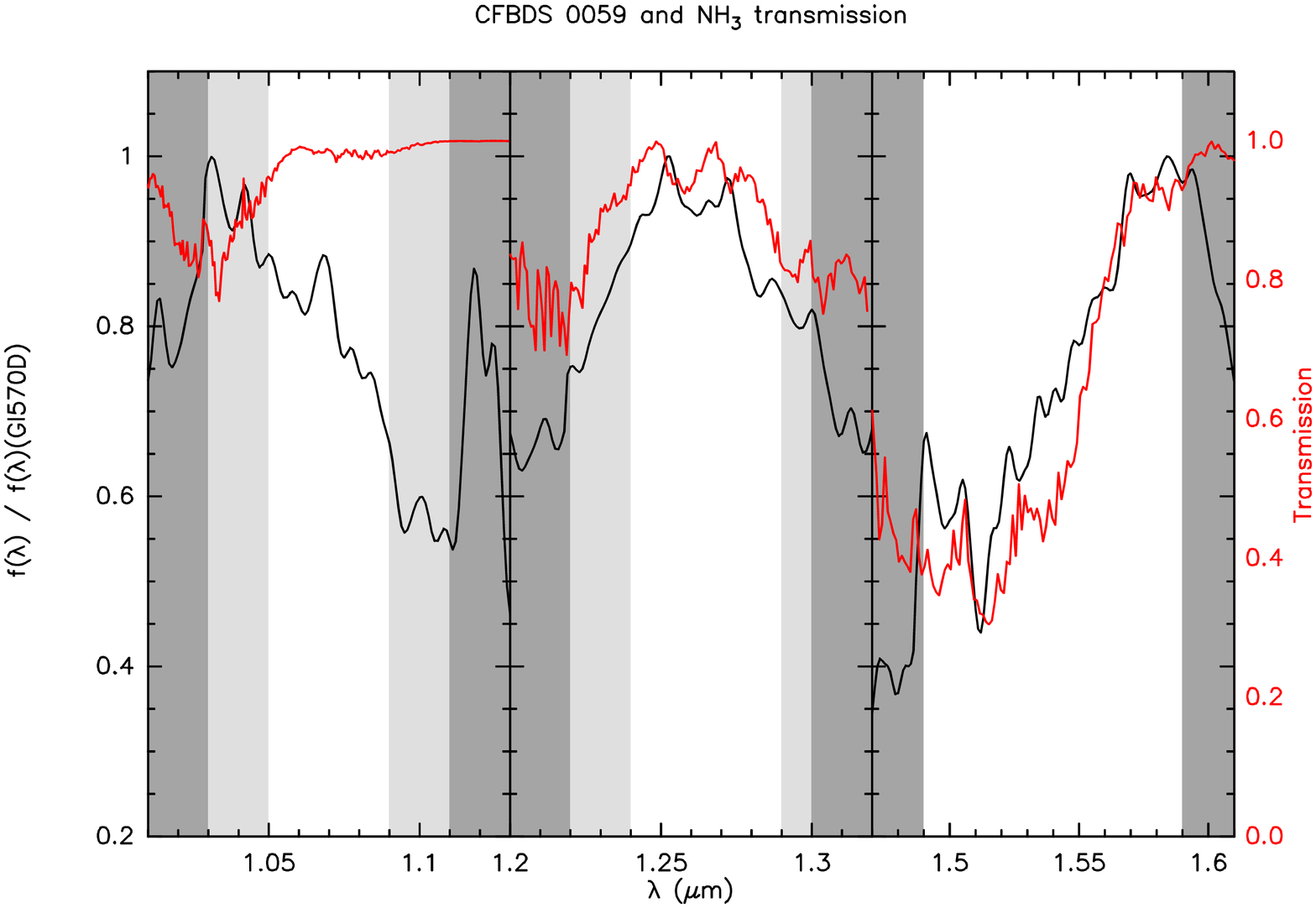} \\
\includegraphics[width=6.6cm,angle=270]{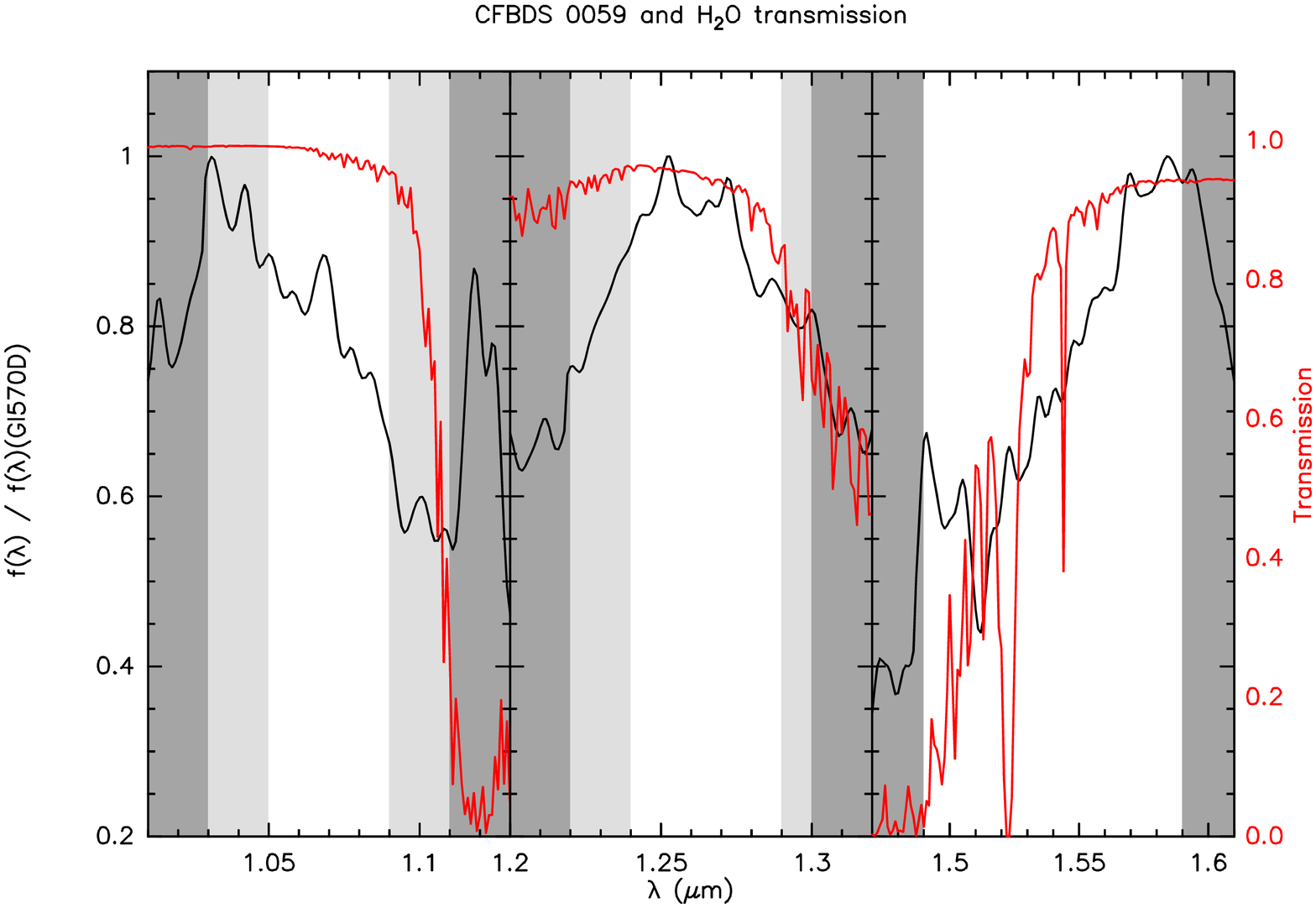}
\end{tabular}
\caption{Flux ratio between CFBDS0059 and Gl~570D (black), together 
with the laboratory room temperature  transmission spectrum of 
NH$_3$ \citep{irwin.1999} (red, top panel) and the 600~K H$_2$O 
 transmission spectrum computed from the HITRAN  molecular 
database (red, bottom panel). The grey bands
mark the parts of the spectrum affected by strong (dark grey) or 
moderate (light grey) CH$_4$ absorption.}
\label{ratio_cfbds}
\end{figure}

\begin{figure}
\begin{tabular}{c}
\includegraphics[width=6.6cm,angle=270]{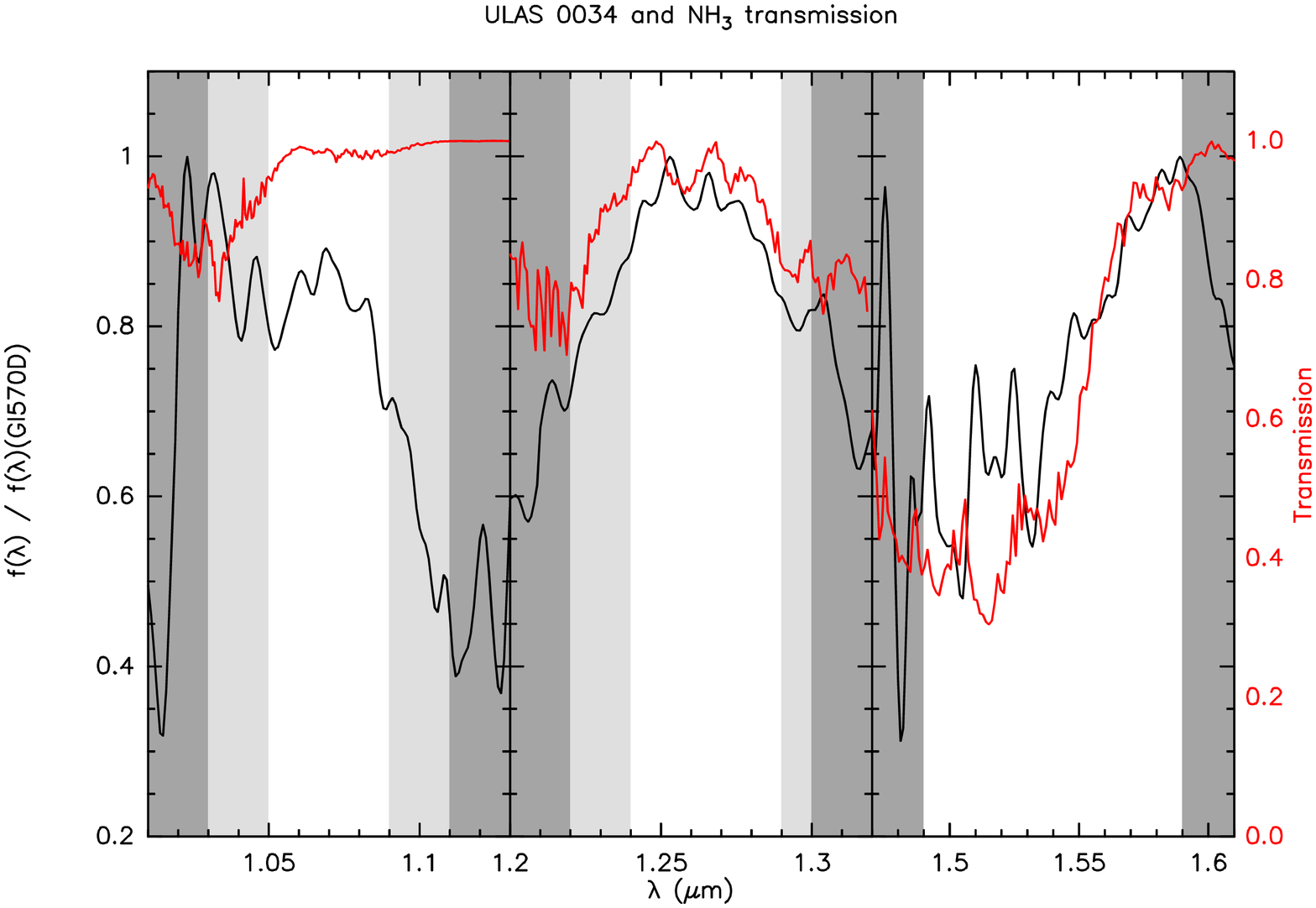} \\
\includegraphics[width=6.6cm,angle=270]{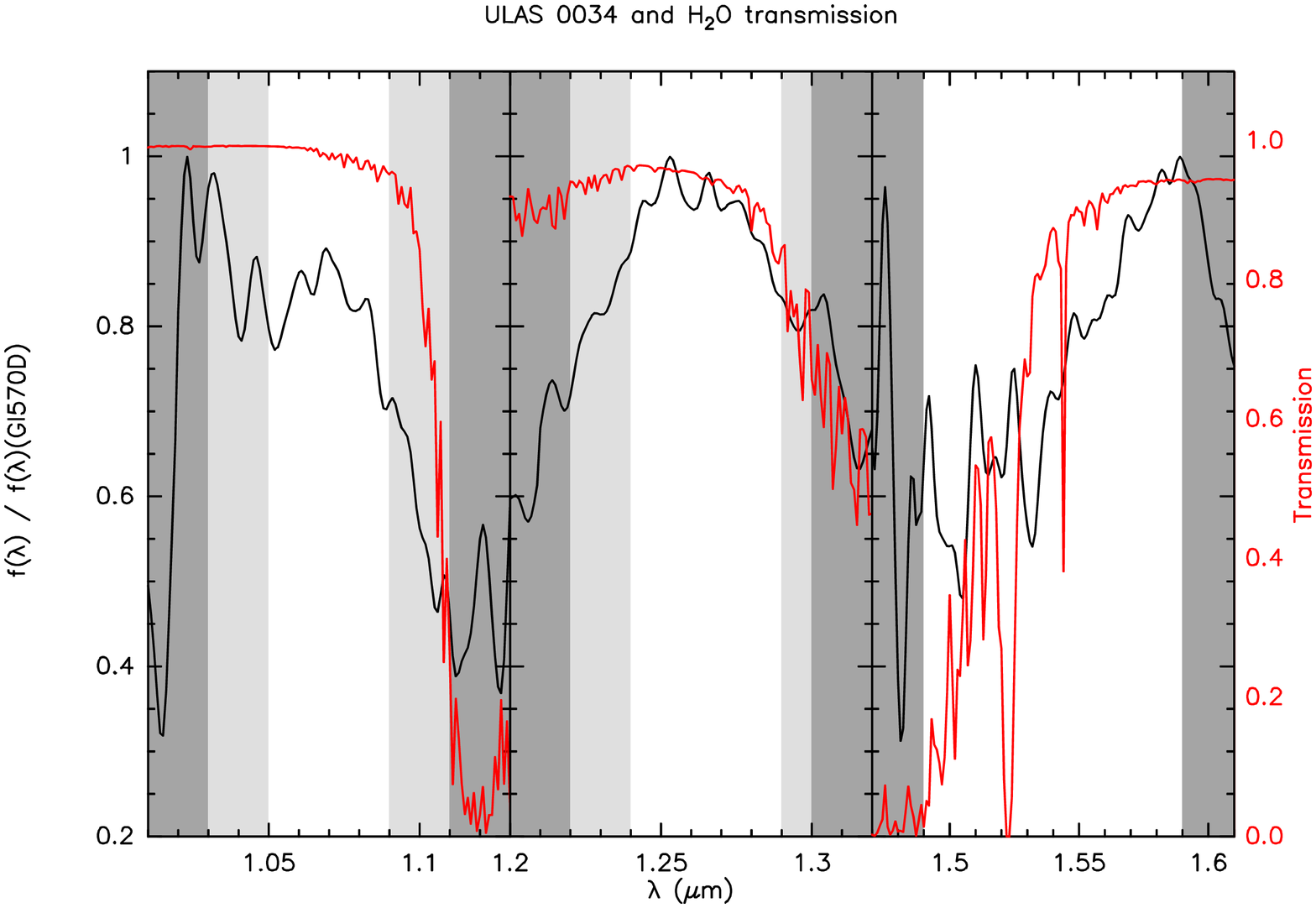}
\end{tabular}
\caption{Flux ratio between ULAS0034 and Gl~570D (black).
The overlays repeat those of Fig~\ref{ratio_cfbds}.}
\label{ratio_ulas}
\end{figure}

 To emphasize the changes in brown dwarfs spectra when their
effective temperature decreases from $\sim$800 to $\sim$600~K, we 
plot in Fig~\ref{ratio_cfbds} the ratio of the spectra of CFBDS0059 and 
Gl~570D. The signal to noise ratio of the resulting $K$-band spectrum 
is too low for detailed analysis, and we therefore focus on the
$Y$, $J$ and $H$ flux peaks. To avoid confusion from changes in 
the temperature-sensitive methane bands, we also mostly ignore 
the parts of the spectrum affected by CH$_4$ absorption bands, hatched 
in dark and light grey for respectively stronger and weaker
bands. Fig.~\ref{ratio_ulas} shows the equivalent plot for ULAS0034,
which is very similar. 

The $H$-band spectrum ratio prominently shows the new absorption band, 
which outside the CH$_4$ band closely matches the 300~K NH$_3$ 
transmission spectrum of \citet{irwin.1999}. Both spectra are 
strongly absorbed between 1.49 and 1.52 $\mu$m and rebound 
from 1.52 to 1.57$\mu$m. Water absorption, by contrast, is a 
poor match to the features of spectrum ratio. The strongest 
water absorption (as computed from the HITRAN molecular 
database for a 600~K temperature) occurs below 1.49$\mu$m, 
at significantly bluer wavelengths than the CFBDS0059 
absorption feature. 

Some weaker but still significant bands of the \citet{irwin.1999}
laboratory ammonia spectrum occur in the $J$ band. Those again match
features of the CFBDS0059/Gl~570D flux ratio, but that agreement is much
less conclusive: water and ammonia absorptions overlap on the red side 
$J$-band peak, and CH$_4$ absorption affects the blue side of the
peak.  A 1.25-1.27 $\mu$m feature is seen on both flux
ratios and on the ammonia transmission, and could be
due to ammonia since it is clear of any strong H$_2$O
absorption band. The slight wavelength shift between
the laboratory and observed features however leaves
that association uncertain.
Detailed synthetic spectra based on fully reliable opacities will
be needed to decide whether NH$_3$ absorption matters in the $J$
band at the effective temperature of CFBDS0059.
The main pattern in the $Y$-band is a blue slope, which reflects
the weaker pressure-broadened K~I wing in the cooler brown dwarf.
The weak 1.03$\mu$m NH$_3$ band is not seen.

Ammonia is easily detected in mid-infrared SPITZER spectra
for all spectral types cooler than $\sim$T2 \citep{Roellig.04,Cushing.06},
though significantly weaker than initially expected because mixing from
lower atmospheric levels
reduces its abundance in the high atmosphere below the local equilibrium 
value \citep{saumon.06}.
Weak near-infrared absorption by ammonia has been tentatively
detected by \citet{saumon.2000} in the T7p dwarfs Gl~229B, but 
CFBDS0059 and ULAS0034 provide the first incontrovertible 
evidence for a strong near-infrared NH$_3$ band in brown 
dwarf spectra.

This conclusion contrasts with \citet{warren.2007} finding
possible but inconclusive evidence for ammonia in ULAS0034.
The main difference between the two analyses is that
\citet{warren.2007} focused on a higher resolution search,
at a necessarily lower signal to noise ratio, for individual
NH$_3$ lines between 1.5 and 1.58$\mu$m. 
We instead looked for the global signature of the absorption
band, which only becomes obvious when looking at the full 
$H$-band spectrum.

\subsection{Spectral type}

Table~\ref{table_indice} lists for CFBDS0059 the spectral indices 
used by the spectral classification scheme of \citet{burgasser.06b}, 
which refines the previous schemes of \citet{geballe.02} and 
\citet{burgasser.02}. These indices would imply a T8 classification, 
identical to that of 2M0415. As discussed above however, the 
near-infrared spectrum of CFBDS0059  demonstrates that it is
over 100~K cooler than 2M0415 and shows clearly different spectral
features. Based on the new indices we present
later, CFBDS0059 should be assigned a later spectral type. The almost
identical \citet{burgasser.06b} indices of the two brown dwarfs
instead reflect those indices measuring H$_2$O and CH$_4$ absorption 
bands which saturate and lose their effective temperature sensitivity at the 
T8 spectral type of 2M0415. Beyond T8 the \citet{burgasser.06b}
classification scheme therefore needs to be extended, with 
new spectral indicators that do not saturate until significantly
later spectral types.

\begin{table*}
\caption{Measured spectral classification indices for CFBDS0059
and ULAS 0034. The first six indices form the base of the \citet{burgasser.06b}
classification scheme, and the table includes the corresponding
spectral classification on that scale. The $W_J$ index is a recent
addition proposed by \citet{warren.2007}.}
\begin{tabular}{lllllll} \hline
Spectral  & numerator  & denominator   & value & spectral              & value & spectral              \\
indice    & wavelength & wavelength    &       & type                  &       & type                  \\
          & \multicolumn{2}{c}{$\mu$m} & \multicolumn{2}{c}{CFBDS0059} & \multicolumn{2}{c}{ULAS0034}  \\ \hline
 H$_2$O-$J$ &  1.14-1.165 &  1.26-1.285  & 0.029$\pm$0.005 &  T8       & 0.012$\pm$0.007  &  $>$T8   \\
 CH$_4$-$J$ &  1.315-1.34 &  1.26-1.285  & 0.165$\pm$0.005 &  T8       & 0.014$\pm$0.009  &  $>$T8   \\
 H$_2$O-$H$ &  1.48-1.52  &  1.56-1.60   & 0.119$\pm$0.008 &  $>$T8    & 0.133$\pm$0.010  &  $>$T8   \\
 CH$_4$-$H$ &  1.635-1.675&  1.56-1.60   & 0.084$\pm$0.002 &  T8       & 0.096$\pm$0.006  &  T8      \\
 CH$_4$-$K$ &  2.215-2.255&  2.08-2.12   & 0.128$\pm$0.037 &  $>$T7    & 0.091$\pm$0.015  &  $>$T7   \\
 $K/J$      &  2.06-2.10  &  1.25-1.29   & 0.101$\pm$0.002 &           & 0.128$\pm$0.003  &          \\
 W$_J$      &  1.18-1.23  &  1.26-1.285  & 0.257$\pm$0.004 &           & 0.262$\pm$0.006  &          \\ \hline
\end{tabular}
\label{table_indice}
\end{table*}

Fully defining this extension is beyond the scope of the present paper,
since two known objects beyond T8 are not enough to explore spectral 
variability, but one can nonetheless start exploring. Since the main 
new feature is NH$_3$ absorption in the blue wings of the $H$-band 
peak, we define a new NH$_3$-$H$ index as

\begin{equation}
NH_3-H = \frac{\int^{1.56}_{1.53}f(\lambda)d\lambda}{\int^{1.60}_{1.57}f(\lambda)d\lambda}.
\end{equation}

 Its numerator and denominator range are plotted in 
Fig~\ref{spectre_comparaison_h2}.
The numerator integrates the flux within the main NH$_3$ band and its 
denominator measures the bulk of the $H$-band peak (we note that in 
cooler objects, the denominator could be affected by some 
NH$_3$ 
absorption; its integration boundaries might thus need to be 
refined after such objects have been discovered). We 
compute this index for Gl~570D, HD3651B, 2M0415, 
ULAS0034 and CFBDS0059  (Table~\ref{table_nh3}), and find  
that it strongly decreases from Gl570D to ULAS0034 and 
CFBDS0059 (which have very similar NH$_3$-$H$).

\begin{table*}
\caption{NH$_3$-$H$ indices for the five brown dwarfs discussed 
in this paper. The spectral classification either $^{(1)}$ follows 
\citet{burgasser.06b} or $^{(2)}$ is derived from the NH$_3$-H and 
W$_J$ indices, as discussed in the text}
\begin{tabular}{llllll} \hline
         & Gl570D & HD3651B &  2M0415 & ULAS0034 & CFBDS0059 \\ \hline
NH$_3$-$H$ & 0.672$\pm$0.008& 0.66$\pm$0.005   &  0.625$\pm$0.003  & 0.516$\pm$0.008 & 0.526$\pm$0.005     \\ 
Spec. Type  & T7.5$^{(1)}$ &  T7.5$^{(1)}$ &  T8$^{(1)}$  & T9 or Y0$^{(2)}$ & T9 or Y0$^{(2)}$ \\ \hline
\end{tabular}
\label{table_nh3}
\end{table*}

\begin{figure}
\includegraphics[width=9.3cm,angle=0]{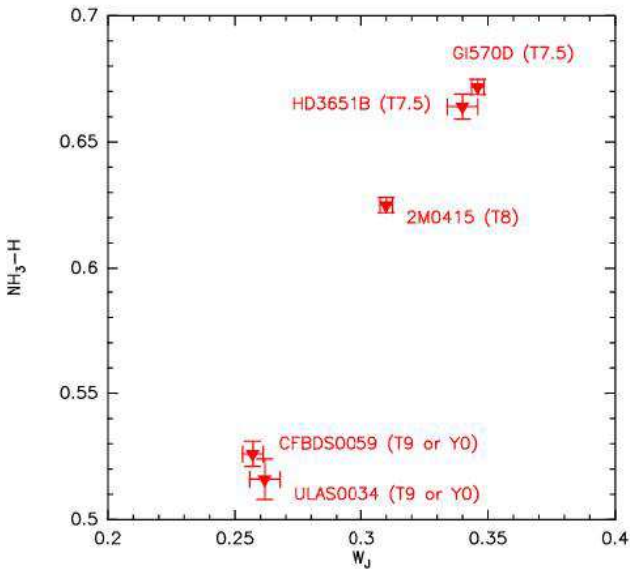}
\caption{NH$_3$-H index versus $W_J$ index. The error bars represent 
the 1$\sigma$ uncertainties of the measured spectral indices.}
\label{wj_nh3}
\end{figure}

Over the limited effective temperature range spanned by Gl570D, HD3651B,
2M0415, ULAS0034 and CFBDS0059, and as far as one can
infer from just 5~examples, the NH$_3$-$H$ and W$_J$ indices 
correlate strongly (Fig.~\ref{wj_nh3}). The numerator 
of W$_J$ is centered at wavelengths where both ammonia 
(Fig.~10 of \citet{leggett.2007}) and CH$_4$ have significant 
opacity, and future modeling work should be able to establish
whether the two indices probe the same molecule or not. Since
the near-infrared spectra of ULAS0034 and CFBDS0059 differ 
significantly more from that of the T8 2M0415 than the latter 
differs from the T7.5 Gl570D (as quantitatively demonstrated 
by Fig.~\ref{wj_nh3}), it is natural to assign a full spectral 
subtype to the interval between the two coolest brown dwarfs and
2M0415. By that reasoning, and if ULAS0034 and CFBDS0059 
are considered as T dwarfs, their spectral type should be 
T9, or perhaps slightly later.

The T spectral class however is quite unlikely to remain the last 
spectral type, since for sufficiently low effective temperatures 
atmospheric models predict major changes in visible and 
near-infrared brown dwarf spectra: NH$_3$ bands are predicted 
to appear in, and eventually to dominate, the near-infrared 
spectrum, the strong pressure-broadened optical lines of Na~I 
and K~I are predicted to disappear as those atomic species get 
incorporated into molecules and solids, and water clouds 
are predicted to form and to largely deplete water from the gas 
phase  \citep{burrows.03,kirkpatrick.05}. 
Since spectral classification, for mostly practical reasons, is 
traditionally based on optical and near-infrared spectra,
such a major transition will justify the introduction of 
a new spectral type, for which the Y letter has long been 
reserved \citep{kirkpatrick.1999,kirkpatrick.2000}. If the
$\lambda~\sim 1.55~\mu$m NH$_3$ band 
 keeps deepening as the effective temperature decreases further,
and eventually becomes a major spectral feature, 
its appearance at $T_\mathrm{eff}\simeq$650~K will become a natural 
transition between the T and Y spectral classes. ULAS0034 
and CFBDS0059 would then be the first Y dwarfs, and the prototypes
for Y0 brown dwarfs, rather than T9. That decision will to some
extent remain a matter of convention, but it must in any case wait
until larger numbers of similarly cool brown dwarfs can
document spectral trends in finer detail, and preferably 
over a wider effective temperature range.

\section{Summary and conclusions}

We have reported the discovery of CFBDS0059, a very cool brown
dwarf, discovered in the CFBDS survey \citep{delorme.2008}. 
Its effective temperature is $\sim$50$\pm$15~K cooler than that of
ULAS0034, most likely making it the coolest brown dwarf known at the present
time.  High spatial resolution imaging establishes than CFBDS0059 
has no similarly bright companion beyond 0.09'', and no companion 
with a contrast under 3.5 magnitude beyond 0.3'' (respectively 
1.2 and 3.9~AU at the 13~pc photometric distance). Its kinematics 
suggest, with significant error bars, a $\sim$4~Gyr age at which 
CFBDS0059 would be a $\sim$30M$_{Jup}$ brown dwarf. The atmospheric
parameters of CFBDS0059 however are compatible with any age from
5~Gyr down to 1~Gyr, for which its mass would be $\sim$15M$_{Jup}$.
A trigonometric parallax measurement together with mid-infrared
photometry and spectroscopy with SPITZER will significantly 
refine its physical parameters, as demonstrated by 
\citet{saumon.07} for slightly warmer brown dwarfs.

We assign absorption
in the blue wing of the $H$-band peaks of both ULAS~0034 and 
CFBDS~0059 to an NH$_3$ band. If that assignment is confirmed,
and if, as we expect, the band deepens at still lower effective 
temperatures, its development would naturally define the scale
of the proposed Y spectral class.  ULAS~0034 and CFBDS~0059 
would then become the prototypes of the Y0 sub-class.

The CFBDS survey has to date identified two brown dwarfs later than
T8, CFBDS0059 and ULAS0034 (which we identified independently
of \citet{warren.2007}, 
\citet{delorme.2008}) in the analysis of approximately 40\% of 
its final 1000~square degree coverage. We therefore expect to 
find another few similarly cool objects, and hopefully one 
significantly cooler one.

 CFBDS0059 and ULAS0034 provide a peek into the atmospheric physics for
conditions that start approching those in giant planets, and the
future discoveries that can be expected from CFBDS, ULAS, and
Pan-STARRS will further close the remaining gap. They also
bring into a sharper light the remaining imperfections of
the atmopsheric models, and emphasize in particular the
importance of more complete opacity data. Our analysis
relies on a room temperature absorption NH$_3$ spectrum,
but higher excitation bands than can be excited at 300~K
must matter in T$_\mathrm{eff}$~=~600~K brown dwarfs. The eventual
identification of ammonia absorption in the $J$ band will
also need complete opacity information for  H$_2$O
and CH$_4$ and full spectral synthesis, since the bands of
the three molecules overlap in that spectral range.

The spectral indices that define the T dwarf spectral class
saturate below 700~K, and new ones will be needed at lower
effective temperatures. We introduce one here, NH$_3$-$H$, which measures
the likely NH$_3$ absorption in the $H$ band. Together with
the $W_J$ index of \citet{warren.2007} and the slope of the blue
side of the $Y$-band peak \citep{leggett.2007}, it will
hopefully define a good effective temperature sequence.
Metallicity and gravity diagnostics are less immediately
apparent, but will need to be identified as well.

\begin{acknowledgements}

We are grateful to our referee, Sandy Leggett, for her
very detailed report and numerous suggestions which significantly 
improved this paper. 
We would like to thank the observers and queue coordinators who carried our
service observations at CFHT (programs 05BC05,06AC20,07BD97) and Gemini-North
(GN-2007A-Q-201, GN-2007B-Q-3). We also thanks the NTT and Keck
Observatory support astronomers for their help during the observations 
which led to these results.
We thank S.~Leggett and S~Warren for providing their spectrum of 
ULAS0034 in a convenient numerical format, and Sandy Leggett
for communicating spectra of 2M0415 and Gl~570D and the room temperature 
absorption spectrum of NH$_3$. We would also like to thank
David Ehrenreich for providing us the computed 600~K water absorption.
Financial support from the "Programme National de Physique Stellaire'' (PNPS) 
of CNRS/INSU, France, is gratefully acknowledged.
MCL acknowledges support for this work from NSF grant AST-0507833 and
an Alfred P. Sloan Research Fellowship.

\end{acknowledgements}

\bibliographystyle{aa}
\bibliography{biblio}

\end{document}